\documentclass[journal,a4paper]{IEEEtran}
\usepackage[pass,a4paper]{geometry}
\usepackage[english]{babel}
\usepackage{graphicx}
\usepackage[pdftex]{xcolor}
\usepackage{amsthm,amsmath,epsfig,amsfonts,dsfont}
\usepackage{cite}
\usepackage[normalem]{ulem}

\newcommand{\bJ}{\boldsymbol{J}}
\newcommand{\bx}{\boldsymbol{x}}
\newcommand{\ba}{\boldsymbol{a}}
\newcommand{\bb}{\boldsymbol{b}}

\newcommand{\bM}{\boldsymbol{M}}
\newcommand{\bm}{\boldsymbol{m}}

\newcommand{\bn}{\boldsymbol{n}}
\newcommand{\bB}{\boldsymbol{B}}

\newcommand{\bD}{\boldsymbol{D}}

\newcommand{\bP}{\boldsymbol{P}}

\newcommand{\bR}{\boldsymbol{R}}
\newcommand{\R}{\mathds{R}}

\newcommand{\bu}{\boldsymbol{u}}
\newcommand{\bv}{\boldsymbol{v}}
\newcommand{\bU}{\boldsymbol{U}}
\newcommand{\bV}{\boldsymbol{V}}

\newcommand{\bY}{\boldsymbol{Y}}
\newcommand{\bZ}{\boldsymbol{Z}}
\newcommand{\bL}{\boldsymbol{L}}
\newcommand{\diag}{\textrm{diag}}

\newcommand{\up}[1]{\text{#1}}

\begin{document}

\title{{
		Colored noise in oscillators.}\\ Phase-amplitude analysis and a method to avoid the It\^o-Stratonovich dilemma}

\author{Michele Bonnin, Fabio L. Traversa, Fabrizio Bonani, \emph{Senior Member, IEEE}\thanks{M. Bonnin and F. Bonani are with the Dipartimento di Elettronica e Telecomunicazioni, Politecnico di Torino, Corso Duca degli Abruzzi 24, 10129 Torino, Italy. F.L. Traversa is with MemComputing Inc., San Diego, CA, 92093-0319, USA.}}


\maketitle

\begin{abstract}
We investigate the effect of time-correlated noise on the phase fluctuations of nonlinear oscillators. The analysis is based on a methodology that transforms a system subject to colored noise, modeled as an Ornstein–Uhlenbeck process, into an equivalent system subject to white Gaussian noise. 
A description in terms of phase and amplitude deviation is given for the transformed system. Using stochastic averaging technique, the equations are reduced to a phase model that can be analyzed to characterize phase noise. We find that phase noise is a drift-diffusion process, with a noise-induced frequency shift related to the variance and to the correlation time of colored noise. The proposed approach improves the accuracy of previous phase reduced models.
\end{abstract}


\section{Introduction}
\label{intro}

Oscillators and phase locked loops (PLLs) are fundamental components of electronic and optical systems. For instance, in digital systems they are used to establish a reference time to synchronize operations. In communication systems, they are used for frequency coding and decoding, and for channel selection. 

Noise sources, both intrinsic and external, are a major nuisance plaguing oscillator and PLL performance. They can be classified as white (frequency independent) fluctuations, such as thermal noise in electrical circuits with resistive elements or shot noise in semiconductor devices, and time-correlated (colored) noise sources. Among the latter, particularly relevant in oscillators based on bipolar transistors and MOSFET devices used as radio frequency sources, we find Lorentzian low-frequency noise and flicker noise. Flicker or $1/f$ fluctuations can be in several cases traced back to the superposition of low-frequency Lorentzian noise sources, that therefore are of paramount importance in assessing oscillator random variations.

The performance and reliability of oscillators depend crucially on noise sources, which deteriorate the oscillator response and are responsible for phase noise and time jitter. Phase noise and time jitter are strictly related concepts, defining oscillator short term frequency instabilities. In particular, phase noise is a frequency domain measure of the oscillator spectral purity, while time jitter describes the time domain accuracy of the oscillator waveforms. Phase noise spreads the bandwidth of the fundamental frequency of oscillators and may produce interference with neighboring channels, thus degrading the whole system performances. Therefore, characterizing phase noise in oscillators is a major problem for practical applications.

Since the seminal work \cite{lesson1966}, linear time invariant models (LTI) have been applied to high-$Q$ resonant and quartz-crystal oscillators. While of great practical importance, such a technique is often too simplistic and fails to capture essential features such as spectral dispersion. Inclusion of linear time variant effects (LTV) can yield more accurate results \cite{hafner1966,kurokawa1968}. 

The most rigorous treatment of phase noise in nonlinear oscillator perhaps dates back to \cite{kaertner1990}, where the author decomposes the oscillator response into phase and magnitude components, and successfully derived a differential equation for the phase deviation. In \cite{hajimiri1998}, the oscillator response is decomposed into orthogonal components, and equations for purely phase and amplitude deviations are derived. Unfortunately, as shown in \cite{demir2000}, using an orthogonal decomposition to separate phase and amplitude deviations leads to inaccurate results. The method proposed in \cite{demir2000}, based on using Floquet vectors to project the response of the noisy oscillator onto a shifted version of the unperturbed response, exploits a linear periodically time varying 
approximation of the oscillator behavior {
leading, ultimately, to a nonlinear phase equation.} The idea that Floquet vectors constitute the ideal basis to decouple phase and amplitude dynamics was confirmed in \cite{coram2001,traversa2011}. Phase domain models based on the ideas introduced in \cite{demir2000}, have been extensively used to derive an analytical stochastic characterization of oscillator phase noise due to both white and colored noise sources \cite{demir2002,gleeson2006,odoherty2007,maffezzoni2007,djurhuus2009,nagashima2014,maffezzoni2014,maffezzoni2016}. 

 In \cite{kaertner1990,demir2000,demir2002}, simplified scalar stochastic differential equations for the phase variable were derived, neglecting all contributions of noise and amplitude fluctuations, i.e small deviations from the limit cycle, beyond the first order. This assumption is well justified for most electronic oscillators, that are typically subject to small noise and exhibit strongly stable limit cycle, thus leading to very accurate phase models. LTV techniques yield reliable and precise predictions for the phase diffusion process, but they fail to capture the frequency shift phenomenon \cite{odoherty2007,bonnin2013,bonnin2016a}. Such a frequency shift is usually negligibly small for strongly stable oscillators, such as those customarily used in electronic systems, but may play a relevant role in autonomous systems from other fields, e.g. system biology and neuroscience \cite{tait2018}.

This paper proposes a novel methodology to analyze phase noise in nonlinear oscillators subject to time-correlated noise, modeled as an Ornstein-Uhlenebeck process. Making use of the method proposed in \cite{givon2004,pavliotis2008} the system with colored noise is first transformed into an equivalent system subject to white Gaussian noise. The advantage is twofold: first the transformation allows to use the whole machinery of stochastic differential equations on a reduced dimensional system. Second, and more important, the transformation avoids the It\^o-Stratonovich dilemma \cite{oksendal2003}. Phase and amplitude deviation equations are derived for the transformed system, and then reduced to a phase model, that describes the oscillator dynamics in terms of the phase variable only. This description is the ideal tool for phase noise analysis, since it gives an approximate yet very accurate description of the phase dynamics. We show that phase noise in nonlinear oscillators is a drift-diffusion process, that is, noise sources not only induce a spread in the oscillator spectrum, but also a shift in the oscillation mean frequency. The frequency shift is related to the variance of the colored noise and to the noise correlation time. Some examples are presented to assess the validity of the approach.  

\section{Modeling}\label{noisy oscillators}

Consider the nonlinear system subject to random noise source
\begin{equation}
\dot \bx_t = \ba(\bx_t) + \bB(\bx_t) \, \eta_t \label{sec1-eq1}
\end{equation}
where $\bx_t : \R \mapsto \R^n$ denotes the state of the system, $\ba:\R^n \mapsto \R^n$ is a smooth vector field that defines the system internal dynamics, $\bB:\R^n \mapsto \R^n$ is a smooth vector valued function 
and $\eta_t : \R \mapsto \R$ is a scalar function describing random fluctuations, both internal and external.

Random fluctuations are often modeled as zero mean, Gaussian distributed white noise. A zero mean Gaussian white noise $\eta_t=\dot W_t$ is characterized by $\langle \dot W_t \rangle=0$ and $\langle \dot W_t \dot W_s \rangle = \delta(t-s)$, where $\delta$ is the Dirac delta function, while $t$ and $s$ denote two time instants. White Gaussian noise is a reasonable approximation in the case where the typical time scales of the underlying deterministic dynamics are much larger than the noise correlation time (quasi-white approximation). Unfortunately, white noise is rarely an accurate model to represent all of the noise sources that induce fluctuations in real electronic devices and systems. As a consequence, a Dirac delta correlation in time is more justified by mathematical convenience than being physically plausible. Real processes have typically finite correlation times, and often $1/f$ power spectra, e.g. \emph{flicker noise}.

A more realistic description of noise in electronic systems is given by an exponentially correlated process. 
Better known as colored noise, it can be modeled as an Ornstein-Uhlenbeck process \cite{gillespie1996}. 
In this case the noise source is modeled according to
\begin{equation}
	\tau \dot \eta_t = - \eta_t + D \dot W_t\label{sec1-eq2}
\end{equation}
where $\tau$ is a parameter proportional to the finite noise correlation time, and $D$ is the diffusion constant. The Ornstein-Uhlenbeck process with deterministic initial state $\eta_0$ is characterized by the expectation value
\begin{equation}
 \langle \eta_t \rangle = \eta_0 \, \exp\left({-\dfrac{t}{\tau}}\right) \label{sec1-eq3}
\end{equation}
and by the correlation function
\begin{equation}
\langle \eta_t \, \eta_s \rangle = \dfrac{D^2}{2 \tau} \exp\left({-\dfrac{|t-s|}{\tau}} \right)\label{sec1-eq4}
\end{equation}
Thus we consider equations \eqref{sec1-eq1} and \eqref{sec1-eq2}, that we rewrite adopting the standard notation of stochastic differential equations (SDEs)
\begin{align}
d \bx_t & = \left[ \ba(\bx_t) + \bB(\bx_t) \eta_t  \right] dt \label{sec1-eq5}\\
\tau d  \eta_t & = - \eta_t  \, dt + D \, dW_t \label{sec1-eq6} 
\end{align}
where $W_t$ is a Wiener process, i.e. the integral of a white noise. 

The SDEs \eqref{sec1-eq5}-\eqref{sec1-eq6} describe a diffusion process with unmodulated (additive) noise. Therefore the equations can be interpreted using any of the two main interpretation schemes, namely It\^o or Stratonovich, obtaining the same solution. In the full space $(\bx,\eta)$ the system is Markovian, that is, future states are completely determined by the current state and the stochastic process shows no memory. Methods for analysis of Markovian systems, based on the Fokker-Planck or Kolmogorov equations, are well developed. However, the practical solution of the equations obtained using these approaches may  become unbearable because of the large number of state variables involved.    

A possible solution strategy amounts to study the system dynamics in a reduced dimension space. However, if we consider the $(\bx)$ space only the system is non Markovian due to the presence of multiplicative noise, as the increments of the state variables depend on the past history of the noise process. Problems arise even in the simpler case of a quasi-white approximation: for $\tau \rightarrow 0$, \eqref{sec1-eq6} shows that the external fluctuations reduce to a white noise $\eta_t dt = D dW_t$.  Substituting this approximation into \eqref{sec1-eq5} yields
\begin{equation}
d \bx_t = \ba(\bx_t) dt + D \, \bB(\bx_t) d W_t  \label{sec1-eq7} 
\end{equation} 
The white noise is now modulated by the state dependent function $\bB(\bx)$ (multiplicative noise), therefore the question arises whether \eqref{sec1-eq7} should be interpreted according to It\^o or Stratonovich.

Applying the procedure presented in \cite{givon2004,pavliotis2008}, in the next section we shall derive a reduced description in the $(\bx)$ space of problem \eqref{sec1-eq5}-\eqref{sec1-eq6} where the SDE system is transformed into an SDE (for the state vector $\bx$ only) subject to white Gaussian noise. The main results are the following
\begin{itemize}
	\item The reduced system holds for small, but not necessarily vanishing correlation time $\tau$.
	\item The reduced system resolves the It\^o-Stratonovich dilemma, that is, we shall derive equivalent SDEs for the two interpretations. By equivalent, we mean two different SDEs, interpreted following different rules, having the same solution. Because the solution is unique, it is just a matter of personal preference to choose one interpretation rather than the other. 
\end{itemize}

\section{White noise approximation}\label{section white noise}

Dividing both sides of \eqref{sec1-eq6} by $\tau$, substituting $\tau = \varepsilon^2$ and introducing $\eta_t = y_t/\varepsilon$, equations \eqref{sec1-eq5}-\eqref{sec1-eq6} become
\begin{align}
d\bx_t & = \left[ \ba(\bx_t) + \dfrac{1}{\varepsilon} \bB(\bx_t) y_t\right] dt \label{sec2-eq1}\\
d y_t & = - \dfrac{1}{\varepsilon^2} \, y_t \,dt + \dfrac{D}{\varepsilon} d W_t \label{sec2-eq2} 
\end{align}
Usually, in electronic systems the correlation time $\tau$ is small compared to the characteristic time constants for the deterministic part of the dynamics. Therefore, we can assume the correlation time $\tau$ small enough such that $\varepsilon\ll 1$. Under this assumption, equations \eqref{sec2-eq1} and \eqref{sec2-eq2} show a time scale separation, since the Ornstein-Uhlenbeck process $y_t$ is one order, in the parameter $\varepsilon$, faster than the state variables $\bx_t$. Notice that a straightforward application of stochastic averaging \cite{khasminskii1968} would lead to inconsistent results. In fact, since asymptotically $\langle y_t \rangle = 0$, the averaged equation would simply coincide with the deterministic (noiseless) system. 

Now, let $u(x,t) = \mathsf{E}[f(\bx_t,t)]$ denote the expected value for a generic, smooth enough function $f(\bx_t,t)$. 
The Kolmogorov backward equation corresponding to \eqref{sec2-eq1}-\eqref{sec2-eq2} takes the form \cite{oksendal2003}
\begin{equation}
\dfrac{\partial u}{\partial t} = \left( \Lambda_0 + \dfrac{1}{\varepsilon} \Lambda_1 + \dfrac{1}{\varepsilon^2} \Lambda_2 \right) u \label{sec2-eq3}
\end{equation}
where 
\begin{align}
\Lambda_0 \, u & =  \dfrac{\partial u}{\partial \bx} \; \ba(\bx) \label{sec2-eq4}\\[1ex]
\Lambda_1 \, u & = y\,\dfrac{\partial u}{\partial \bx} \; \bB(\bx)  \label{sec2-eq5} \\[1ex]
\Lambda_2 \, u & = -  y \, \dfrac{\partial u}{\partial y} + \dfrac{D^2}{2} \, \dfrac{\partial^2 u}{\partial y^2} \label{sec2-eq6}
\end{align}
and $\partial u/\partial \bx$ is a row vector denoting the gradient of the scalar function $u$ with respect to the vector $\bx$ (thus $\partial u/\partial \bx \; \ba(\bx)$ is the scalar product of the two vectors).
We look for a solution of \eqref{sec2-eq3} in the form of a power series expansion $u = u_0 + \varepsilon u_1 + \varepsilon^2 u_2 + \ldots$\\
Introducing this ansatz into \eqref{sec2-eq3} and equating the coefficients of the same powers of $\varepsilon$ yields the hierarchy of equations
\begin{align}
\varepsilon^{-2}: \qquad \Lambda_2 \, u_0 & = 0 \label{sec2-eq7} \\[1ex]
\varepsilon^{-1}: \qquad\Lambda_2 \, u_1 & = - \Lambda_1 u_0 \label{sec2-eq8} \\[1ex]
\varepsilon^{0}: \qquad\Lambda_2 \, u_2 & = \phantom{-} \dfrac{\partial u_0}{\partial t} - \Lambda_0 u_0 - \Lambda_1 u_1 \label{sec2-eq9} \\[1ex]
\nonumber \vdots &   
\end{align}
The first equation in the hierarchy implies that $u_0$ does not depend on $y$, so that $u_0 = u_0(\bx,t)$. 

The other equations are of the type $\Lambda_2 u_n = b_n$. According to Fredholm alternative theorem, these equations are solvable provided that a function $\psi$ exists such that: (1) $\Lambda_2^* \, \psi = 0$, where $\Lambda_2^*$ is the conjugate operator of $\Lambda_2$, and  (2) each $b_n$ satisfies $(b_n,\psi)=0$, where $(\cdot,\cdot)$ denotes the inner product in the $L^2$ Banach space. Taking into account \eqref{sec2-eq6}, $\Lambda_2^* \, \psi =0$ implies that $\psi$ is the stationary distribution of the Ornstein-Uhlenbeck process \eqref{sec2-eq2}, thus \cite{gardiner1985}
\begin{equation}
\psi(y)=p_\up{st}(y) = \sqrt{\dfrac{1}{\pi D^2}} \; \exp\left(-\dfrac{y^2}{D^2}\right) \label{sec2-eq10}
\end{equation} 
As a consequence, condition $(b_n,\psi)=0$ amounts to require that each term $b_n$ averages to zero with respect to $y$
\begin{equation}
(b_n,\psi) = \langle b_n \rangle_y = \int_{\R} b_n \, p_\up{st}(y) \, dy = 0 \label{sec2-eq11}
\end{equation}
Taking into account that $\langle y \rangle_y = 0$, it is straightforward to verify that $\langle \Lambda_1 u_0 \rangle_y = 0$. Thus equation \eqref{sec2-eq8} is solvable, and direct substitution shows that
\begin{equation}
u_1(\bx,y,t) = y \dfrac{\partial u_0(\bx,t)}{\partial \bx} \; \bB(\bx)= -\Lambda_2^{-1} \Lambda_1 u_0 \label{sec2-eq13}
\end{equation}

Similarly, equation \eqref{sec2-eq9} is solvable if
\begin{equation}
\left\langle \dfrac{\partial u_0}{\partial t} \right\rangle_y - \left\langle \Lambda_0 u_0 \right\rangle_y + \left\langle \Lambda_1 \Lambda_2^{-1} \Lambda_1 u_0 \right\rangle_y = 0 \label{sec2-eq14}
\end{equation}
The three averages on the left hand side can be expressed as
\begin{align}
&\int_{\R} \dfrac{\partial u_0(\bx,t)}{\partial t} \, p_\up{st}(y) \, dy  =   \dfrac{\partial u_0(\bx,t)}{\partial t} \label{sec2-eq15}\\[1ex]
& \int_{\R} \Lambda_0 u_0 \, p_\up{st}(y) \, dy = \dfrac{\partial u_0(\bx,t)}{\partial \bx} \; \ba(\bx) \label{sec2-eq16} \\[1ex] 
&\nonumber \int_{\R} \Lambda_1 \Lambda_2^{-1} \Lambda_1  u_0 \, p_\up{st}(y) \, dy = - \dfrac{D^2}{2} \bigg[ \dfrac{\partial u_0(\bx,t)}{\partial \bx} \, \dfrac{\partial \bB(\bx)}{\partial \bx} \, \bB(\bx)\\[1ex]
& \hspace{35mm} + \bB^T(\bx) \, \dfrac{\partial^2 u_0(\bx,t)}{\partial \bx^2} \, \bB(\bx) \bigg]  \label{sec2-eq17}
\end{align}	
where $\partial \bB(\bx)/\partial \bx$ is the Jacobian matrix of the vector function $\bB$ with respect to the $\bx$ variables with elements $(\partial \bB(\bx)/\partial \bx)_{ij} = \partial B_i(\bx)/\partial x_j$,  while $\partial^2 u_0(\bx,t)/\partial \bx^2$ is the Hessian matrix of the scalar function $u_0$ with respect to $\bx$ with elements $(\partial^2 u_0(\bx,t)/\partial \bx^2)_{ij} = \partial^2 u_0(\bx,t)/(\partial x_i \partial x_j)$. Furthermore, to solve the last integral we used the fact that, from \eqref{sec2-eq10}, $\langle y^2 \rangle_y = D^2/2$.

Substituting equations \eqref{sec2-eq15}-\eqref{sec2-eq17} into \eqref{sec2-eq14} yields the Kolmogorov backward equation
\begin{align}
\nonumber \dfrac{\partial u_0(\bx,t)}{\partial t} = &  \dfrac{\partial u_0(\bx,t)}{\partial \bx} \; \ba(\bx) + \dfrac{D^2}{2} \bigg[ \dfrac{\partial u_0(\bx,t)}{\partial \bx} \dfrac{\partial \bB(\bx)}{\partial \bx} \bB(\bx) \\[1ex] 
& + \bB^T(\bx) \, \dfrac{\partial^2 u_0(\bx,t)}{\partial \bx^2} \, \bB(\bx) \bigg]\label{sec2-eq18}
\end{align}
%

The corresponding Stratonovich SDE is (we use the symbol $\circ$ to denote Stratonovich stochastic integral)
\begin{equation}
d \bx_t = \ba(\bx_t) dt + D \, \bB(\bx_t) \circ dW_t \label{sec2-eq19}
\end{equation} 
while the equivalent It\^o SDE is
\begin{equation}
d \bx_t = \left[ \ba(\bx_t) + \dfrac{D^2}{2} \dfrac{\partial \bB(\bx_t)}{\partial \bx} \, \bB(\bx_t) \right] dt + D \, \bB(\bx_t) \, dW_t \label{sec2-eq20}
\end{equation} 
By equivalent we mean that the SDEs \eqref{sec2-eq19} and \eqref{sec2-eq20} are interpreted according to different rules but they have the same solution. 
The Stratonovich SDE \eqref{sec2-eq19} can be transformed into the It\^o SDE \eqref{sec2-eq20} (and vice versa) by addition (respectively, subtraction) of the Wong-Zakai drift correction term \cite{oksendal2003}. The solution of \eqref{sec2-eq19} and \eqref{sec2-eq20} is also a weak solution for the original system with colored noise \eqref{sec1-eq5}-\eqref{sec1-eq6}. Weak means that the two solutions for a specific realization of the the noise process $d W_t$ are different in details, but they converge in probability, i.e. they have the same probability density function and therefore the same statistical properties.

Whether to use the Stratonovich SDE \eqref{sec2-eq19} or the equivalent It\^o SDE \eqref{sec2-eq20} is at this point a matter of personal taste. As a rule of thumb, Stratonovich interpretation may be better suited for algebraic manipulations, since traditional calculus rules apply. By contrast, It\^o interpretation requires a whole new set of calculus rules, known as It\^o calculus \cite{oksendal2003}, but it is more suitable for numerical simulations and for the calculation of expected quantities.

\section{Phase-amplitude equations for nonlinear oscillators with colored noise}\label{section phase-amplitude}

Let us now consider the case where equations \eqref{sec1-eq5}-\eqref{sec1-eq6} describe a nonlinear oscillator subject to colored noise. In the absence of random fluctuations, equation \eqref{sec1-eq5} reduces to the autonomous ordinary differential equation (ODE)
\begin{equation}
\dfrac{d \bx}{dt} = \ba(\bx) \label{sec3-eq1}
\end{equation} 
We assume that \eqref{sec3-eq1} admits of an asymptotically stable $T$-periodic solution $\bx_s(t)$, represented by a limit cycle in its state space. 

We shall derive an equivalent description of system \eqref{sec2-eq19} (or \eqref{sec2-eq20}) in terms of phase and amplitude deviation variables, analogous to the one derived in \cite{Traversa2015,bonnin2016a,bonnin2016b}. The phase function used in our description coincides locally, in the neighborhood of the limit cycle, with the asymptotic phase defined in \cite{demir2000,djurhuus2009,mauroy2018}. As a second step, we shall derive a phase reduced model, that describes the oscillator dynamics in terms of the phase variable alone. 

For our purpose it is more convenient to work with the It\^o SDE \eqref{sec2-eq20}. The reason to prefer the It\^o over the Stratonovich interpretation is that It\^o integrals are adapted processes, i.e. state variables and the noise increment are independent. By contrast, in the Stratonovich interpretation state variables and noise increments are correlated, a property known as ``anticipating nature'' or ``look in the future property'' of the Stratonovich integral. The far reaching consequence is that when one tries to describe the dynamics by using only a subset of state variables, an additional piece of information is lost, represented by the correlation between  eliminated variables and noise increments \cite{bonnin2013}.

To make the paper self-contained, we introduce some notation. We consider a set of time dependent vectors $\{\bu_1(t),\ldots,\bu_n(t)\}$, forming a basis for $\R^n$, for all $t$. These basis vectors can be conveniently constructed as follows: the vector $\bu_1(t)$ is chosen as the unit vector tangent to the limit cycle at any $t$
\begin{equation}
\bu_1(t) = \dfrac{\ba(\bx_s(t))}{|\ba(\bx_s(t))|} \label{sec3-eq2}
\end{equation}
The remaining $n-1$ vectors $\bu_2(t),\ldots,\bu_n(t)$ can be chosen as the Floquet vectors (apart from the limit cycle tangent $\bu_1(t)$) of the linearized variational equation \cite{Traversa2015,bonnin2016a,bonnin2016b}
\begin{equation}
\dfrac{d \widetilde\bx(t)}{d t} = \bJ_{\ba}(t) \, \widetilde \bx(t) \label{sec3-eq3}
\end{equation}
where $\bJ_{\ba}(t) = \partial \ba/\partial \bx$ is the Jacobian matrix of the vector function $\ba(\bx(t))$ evaluated on the limit cycle $\bx_s(t)$. Thus the vectors $\{\bu_1(t),\ldots,\bu_n(t)\}$ are independent, although in general they are not orthogonal. We construct the matrix $\bU(t) = [\bu_1(t),\ldots,\bu_n(t)]$, and we define the reciprocal vectors $\bv_1^T(t),\ldots,\bv_n^T(t)$ to be the rows of the inverse matrix $\bV(t) = \bU^{-1}(t)$. Thus $\{\bv_1(t),\ldots,\bv_n(t)\}$ also span $\R^n$ and the bi-orthogonality condition $\bv_i^T \bu_j = \bu_i^T \bv_j = \delta_{ij}$ for all $t$, holds. Finally we introduce matrices $\bY(t) = [\bu_2(t),\ldots,\bu_n(t)]$, $\bZ(t) = [\bv_2(t),\ldots,\bv_n(t)]$, and the magnitude (in the $L^2$ norm) of the vector field evaluated on the limit cycle, $r(t) = |\ba(\bx_s(t))|$.

Following \cite{Traversa2015,bonnin2016a,bonnin2016b}, we decompose the solution of \eqref{sec2-eq20} into two components
\begin{equation}
\bx_t = \bx_s(\theta_t) + \bY(\theta_t) \bR_t \label{sec3-eq4} 
\end{equation}
The first component $\bx_s(\theta_t)$ represents the projection of the stochastic process $\bx_t$ onto the limit cycle, evaluated at an unknown time instant $\theta_t$. The second component $\bY(\theta_t) \bR_t$ represents the distance between the solution and the limit cycle, measured along the directions spanned by the vectors $\bv_2,\ldots,\bv_n$ at the random time $\theta_t$. Because $\bx_t$ is a stochastic process, both $\theta:\R\mapsto\R$ and $\bR:\R\mapsto\R^{n-1}$ are stochastic processes as well. It\^o equations defining the time evolution of these stochastic processes can be found following the procedure given in \cite[Theorem 3.1]{bonnin2016a} and \cite[Theorem 1]{bonnin2016b}, obtaining
\begin{align}
\nonumber d \theta_t = &  \left[ 1 + a_{\theta}(\theta_t,\bR_t) + \hat a_{\theta}(\theta_t,\bR_t) + b_{\theta}(\theta_t,\bR_t) \right] dt \\
&  + B_{\theta}(\theta_t,\bR_t) \, dW_t \label{sec3-eq5}\\[1ex]
d \bR_t = & \left[ \bL(\theta_t) \bR_t + \ba_{\bR}(\theta_t,\bR_t)+ \right. \nonumber\\[1ex] 
&\left. + \hat \ba_{\bR}(\theta_t,\bR_t) + \bb_{\bR}(\theta_t,\bR_t) \right] dt+ \nonumber\\[1ex]
& + \bB_{\bR}(\theta_t,\bR_t) \, dW_t \label{sec3-eq6}
\end{align}
where (the $'$ sign denotes the derivative with respect to $\theta$)
\begin{align}
a_{\theta}(\theta,\bR) = & \kappa \, \bv_1^T \left[ \ba(\bx_s+\bY \bR) - \ba(\bx_s) - \bY' \bR \right] \label{sec3-eq7} \\[1ex]
\nonumber \hat a_{\theta}(\theta,\bR) = & -\kappa \, \bv_1^T \bigg[\bY' \bB_{\bR}(\theta,\bR)  B_{\theta}(\theta,\bR) \\
&  + \dfrac{1}{2}  B_{\theta}^2(\theta,\bR) (\bx_s'' + \bY'' \bR) \bigg] \label{sec3-eq8} \\[1ex]
b_{\theta}(\theta,\bR) = & \dfrac{D^2}{2} \, \kappa \,  \bv_1^T \dfrac{\partial \bB(\bx_s+\bY \bR)}{\partial \bx} \bB(\bx_s+\bY\bR) \label{sec3-eq9} \\[1ex]
B_{\theta}(\theta,\bR) = & D \, \kappa \, \bv_1^T \bB(\bx_s + \bY \bR) \label{sec3-eq10}
\end{align}
\begin{align}
\bL(\theta) = & - \bZ^T \, \bY' \label{sec3-eq11} \\[1ex]
\ba_{\bR}(\theta,\bR) = & \bZ^T \left[ \ba(\bx_s + \bY \bR) - \bY' \bR \, a_{\theta}(\theta,\bR)\right] \label{sec3-eq12} \\[1ex]
\nonumber \hat \ba_{\bR}(\theta,\bR) = & - \bZ^T \bigg[ \bY' \bR \, \hat a_{\theta}(\theta,\bR) + \bY' \bB_{\bR}(\theta,\bR)  B_{\theta}(\theta,\bR) \\
&  + \dfrac{1}{2}  B_{\theta}^2(\theta,\bR) (\bx_s'' + \bY'' \bR) \bigg] \label{sec3-eq13} \\[1ex]
\nonumber \bb_{\bR}(\theta,\bR) = & -\bZ^T \bY' \bR \, b_{\theta}(\theta,\bR) \\
& + \dfrac{D^2}{2} \, \bZ^T \dfrac{\partial \bB(\bx_s+\bY \bR)}{\partial \bx} \bB(\bx_s+\bY\bR) \label{sec3-eq14} \\[1ex]
\bB_{\bR}(\theta,\bR) = & - \bZ^T \bY' \bR \, B_{\theta}(\theta,\bR) + D \, \bZ^T \bB(\bx_s + \bY \bR) \label{sec3-eq15}
\end{align}
and 
\begin{equation}
\kappa = \left(r+\bv_1^T \bY' \bR \right)^{-1} \label{sec3-eq16}
\end{equation}

The SDEs \eqref{sec3-eq5}-\eqref{sec3-eq6} describe phase  noise in nonlinear oscillators with colored noise as a drift-diffusion process. The responsibility of random fluctuations to phase diffusion does not come as a surprise, since, contrary to the amplitude, phase deviation does not have a self-limiting mechanism. Phase deviations are not damped, and may eventually grow unbounded as time passes. However, noise is also responsible for phase drift, that is, it produces a shift in the position of the peaks of the oscillator frequency spectrum. Because of the nonlinear response of the oscillator, random forces applied at a certain angle are amplified, while other are reduced. This results in a net, non null contribution to the expected angular frequency. Noise induced frequency shift is also observed in nonlinear oscillators subject to white Gaussian noise \cite{bonnin2013,bonnin2016a,bonnin2016b}, but in presence of colored noise there is an additional shift contribution, represented by the term $b_{\theta}$, that can be ascribed to the finite correlation time of the noise source. 

\section{Phase equation}\label{phase equation}
The phase and amplitude deviation SDEs \eqref{sec3-eq5}-\eqref{sec3-eq6} have the same solution as the Stratonovich SDE \eqref{sec2-eq19} or the It\^o SDE \eqref{sec2-eq20}, that in turn are characterized by the same statistical properties of the solution of the SDEs \eqref{sec1-eq5}-\eqref{sec1-eq6}. Since \eqref{sec3-eq5}-\eqref{sec3-eq6} are  exact, they are not easier to solve than the white noise approximated SDE \eqref{sec2-eq20}. However, they can be used to derive a phase reduced model \cite{Traversa2015,nakao2016,freitas2018} that in turn can form the basis to find useful, albeit approximate, results. The main advantage of a phase reduced model is that  methods for Markovian systems, e.g. Fokker-Planck and Kolmogorov equations, can be (comparatively) easily applied, and the obtained  equations can be more easily solved, being one dimensional (single variable). 

To derive a simplified phase equation, we exploit a stochastic averaging technique. First we observe that if the Floquet basis is used as vectors $\bu_i,\,\bv_j$, then 
\begin{equation}
a_{\theta}(\theta,\bR) = \mathcal{O}(\bR^2) \label{sec3-eq17} 
\end{equation}
where $\mathcal{O}(\bR^2)$ denotes terms quadratic in the amplitude deviation components, see \cite[Theorem 3]{bonnin2016b} for a proof. Then in the neighborhood of the limit cycle the ``deterministic'' frequency shift $a_{\theta}$ becomes negligibly small, and all points rotate with a uniform angular frequency. In many of the practical applications noise perturbations are small if compared to deterministic effects, that is, $\hat a_{\theta}(\theta,\bR)$, $b_{\theta}(\theta,\bR)$ and $B_{\theta}(\theta,\bR)$ can be considered as perturbation terms.  They either include some explicit small parameter, or the condition $D \ll 1$ holds. As a consequence, the stationary distribution for the angle is expected to remain close to the uniform distribution $p_\up{st}(\theta) = 1/(2\pi)$, that describes the phase diffusion process in a nonlinear oscillator with uniform angular frequency \cite{bonnin2016a,bonnin2016b}.

Similar considerations can be made for the amplitude deviation SDE \eqref{sec3-eq6}. It can be shown that (see \cite[Theorem 3]{bonnin2016b})
\begin{equation} 
\bL(\theta) \bR + \ba_{\bR}(\theta,\bR) = \bD \bR + \mathcal{O}(\bR^2) \label{sec3-eq18}
\end{equation}
where $\bD = \diag[\nu_2,\ldots,\nu_n]$ is a diagonal matrix whose entries are the Floquet characteristic exponents, with the exception of the structural one $\nu_1 = 0$. The amplitude deviation dynamics is the balance of two competing forces: random fluctuations drive the system out of the limit cycle, while the asymptotic stability of the limit cycle implies that the system is continuously pushed toward the periodic orbit. Electronic systems are usually strongly stable, meaning that $\up{Re}\{\nu_i\} \ll 0$ for all $i=2,\ldots,n$, and as a consequence amplitude fluctuations remain confined to a small neighborhood of the limit cycle. Thus we can linearize the amplitude deviation SDE around the noiseless solution $\bR=0$, and after averaging with respect to the phase stationary distribution $p_\up{st}(\theta)=1/(2\pi)$ we obtain
\begin{equation}
d \bR_t = \left( \bM \bR_t + \bm \right) dt +  \bn \,  dW_t   \label{sec3-eq19}
\end{equation}
where (as usual $\partial \hat \ba_{\bR}/\partial \bR$, $\partial \bb_{\bR}/\partial \bR$ and $ \partial \bB_{\bR}/\partial \bR$ are the Jacobian matrices with respect to $\bR$)
\begin{align}
\boldsymbol M = & \bD + \left \langle \dfrac{\partial \hat \ba_{\bR}}{\partial \bR} \right \rangle_{\theta} + \left \langle \dfrac{\partial \bb_{\bR}}{\partial \bR} \right \rangle_{\theta}  \label{sec3-eq20} \\[1ex]
\boldsymbol m = & \left \langle \hat \ba_{\bR}  \right \rangle_{\theta} +  \left \langle \bb_{\bR} \right \rangle_{\theta} \label{sec3-eq21} \\[1ex]
\boldsymbol n = &\left \langle  \bB_{\bR} \right \rangle_{\theta} \label{sec3-eq23}
\end{align}
and 
\begin{equation}
\left \langle f(\theta) \right \rangle_{\theta} = \int_0^{2 \pi} f(\theta) \, p_\up{st}(\theta) \, d \theta = \dfrac{1}{2\pi} \int_0^{2\pi} f(\theta) \, d \theta \label{sec3-eq24}
\end{equation}
In general, the solution of the linear SDE \eqref{sec3-eq19} is not a Gaussian process, but the vector of the expected values $\boldsymbol \mu(t) = \langle \bR_t \rangle $ and the matrix of second moments $\bP(t) = \langle \bR_t \, \bR_t^T \rangle$ can be found solving the linear ODE \cite{kloeden1992}
\begin{align}
\dfrac{d \boldsymbol \mu}{dt} = & \bM \boldsymbol \mu + \bm \label{sec3-eq25} \\[1ex]
\dfrac{d \bP}{dt} = & \bM  \bP + \bP \bM^\up{T} +  \bm \boldsymbol \mu^\up{T} + \boldsymbol \mu \bm^\up{T}  + \bn \bn^\up{T}  \label{sec3-eq26}
\end{align}

Finally, the first and second moment are used to obtain a phase reduced equation. Expanding the terms of the phase SDE \eqref{sec3-eq5} in Taylor series around $\bR=0$, and averaging with respect to the amplitude deviation, yields
\begin{align}
\nonumber d \theta = & \bigg[ 1 + \hat a_{\theta} + b_{\theta}  + \sum_{i} \bigg( \dfrac{\partial \hat a_{\theta}}{\partial R_i} + \dfrac{\partial b_{\theta}}{\partial R_i}  \bigg) \mu_i \\
\nonumber & + \dfrac{1}{2} \sum_{i,j} \bigg( \dfrac{\partial^2 a_{\theta}}{\partial R_i \partial R_j} + \dfrac{\partial^2 \hat a_{\theta}}{\partial R_i \partial R_j} + \dfrac{\partial^2 b_{\theta}}{\partial R_i \partial R_j} \bigg) P_{ij} \bigg] dt \\[1ex]
& + \bigg( B_{\theta} + \sum_i \dfrac{\partial B_{\theta}}{\partial R_i} \mu_i + \dfrac{1}{2} \sum_{i,j} \dfrac{\partial^2 B_{\theta}}{\partial R_i \partial R_j} P_{ij} \bigg) dW_t \label{sec3-eq27}
\end{align}  
where the functions $a_{\theta}, \, \hat a_{\theta}, \, b_{\theta}, \, B_{\theta}$ and their derivatives are evaluated at $(\theta,0)$, and $R_i$ denotes the $i$-th component of vector $\bR$.
	
 We compare the phase equation \eqref{sec3-eq27} with the analogous equations obtained for a nonlinear oscillator subject to white Gaussian noise \cite{bonnin2016a,bonnin2016b}. Apart from 1 that represents the oscillator's normalized angular frequency, the terms in the first two rows describe a frequency shift, whereas the terms in the last row describe a diffusion. $\hat a_{\theta}$ and its derivatives resolve the correlation between the phase and noise increments. These terms were already discussed for a nonlinear oscillator subject to white Gaussian noise \cite{bonnin2016a,bonnin2016b}. By contrast, $b_{\theta}$ and its derivatives describe the different action that colored noise exerts on the phase with respect to white noise only, due to the non null noise correlation time. 
	
It is worth noticing that in the weak noise limit, if higher order contributions of amplitude fluctuations and the correlation resolving term are neglected (implying  $\mu_i = 0, \, P_{ij}=0$ and $\hat a_{\theta}(\theta,\bR)=0$, respectively), then the simplified phase equation is obtained 
\begin{equation}
d \theta = \left[1 + b_{\theta}(\theta,0) \right] dt + B_{\theta}(\theta,0) dW_t \label{sec3-eq30}
\end{equation}
Eq. \eqref{sec3-eq30} is the equivalent of the phase equations derived in \cite{kaertner1990,demir2000} for the case of colored noise, where $b_{\theta}(\theta,0)$ is a zero order approximation of the frequency shift effect produced by the finite noise correlation time.

In order to compare our results against previous literature on phase noise in oscillators subject to colored noise sources, we consider here the approach developed in \cite{demir2002} where higher order contributions of amplitude fluctuations are neglected. For a strongly stable limit cycle, fluctuations are expected to keep the trajectory in a small neighborhood of the limit cycle, so that amplitude noise plays no influence on the phase dynamics. The noisy solution is then approximated as a time shifted version of the noiseless limit cycle $\bx_t = \bx_S(\theta_t)$, and a phase equation is readily derived (see \cite{demir2002}, eq. (3)\footnote{The division by $r(\theta)$, absent in \cite{demir2002}, is a normalization required to guarantee that the oscillator's free running frequency is equal to one.}). For our system \eqref{sec1-eq5}, \eqref{sec1-eq6} the phase model in \cite{demir2002} reads 
\begin{align}
 d \theta = & \left[ 1 + \dfrac{\bv_1^T(\theta) \, \bB(\bx_S(\theta))}{r(\theta)} \, \eta_t \right] dt \label{sec3-eq28} \\[1ex]
 \tau d \eta_t = & - \eta_t \, dt + D \, dW_t \label{sec3-eq29}
\end{align}
Notice that this phase equation is further approximated in \cite{demir2002} to derive the noise spectra.

Fully neglecting the amplitude fluctuations may be a reasonable approximation for strongly stable oscillators, however a more detailed analysis shows that in some cases the amplitude noise impacts on the cycle frequency inducing a non-negligible shift \cite{odoherty2007,bonnin2013,bonnin2016a}, especially for some autonomous systems exploited in computational biology and neuroscience \cite{tait2018}.

\section{Examples}\label{section example}

\subsection{Stuart-Landau oscillator with colored noise}

As a first example we consider a Stuart-Landau oscillator with colored noise. The reason to choose such a simple system is twofold. First, most of the analysis can be made analytically, making the example useful to illustrate the theory and techniques described in the previous sections. Second, because many of the equations admit of an exact solution, the example permits to assess the accuracy of exploited approximations. 

The state equations are the following
\begin{equation}
\begin{array}{rl}
d \phi = & \left(\alpha - \beta \rho^2 + \rho \, \eta_t \right) dt \\[1ex]
d \rho = &  (\rho - \rho^3 + \rho^2 \eta_t) dt  \\[1ex]
\tau d \eta_t  = & - \eta_t \, dt + D \, dW_t  
\end{array}\label{sec4-eq1}
\end{equation}
where $\alpha$ and $\beta$ are real parameters that define the oscillator free running frequency. 

Applying the methodology described in section \ref{section white noise} we obtain the following SDE with modulated white Gaussian noise
\begin{equation}
\begin{array}{rl}
d \phi = & \left[ \alpha + \left(\dfrac{D^2}{2} - \beta \right) \rho^2 \right] dt + D \, \rho \, dW_t \\[2ex]
d \rho = & \left[ \rho + \left(D^2-1 \right) \rho^3 \right] dt + D \, \rho^2 \, dW_t 
\end{array}\label{sec4-eq2}
\end{equation}
We can now take advantage of the particularly simple structure of the Stuart-Landau system. Because the SDE for the amplitude is independent on the phase, the Fokker-Planck equation for the amplitude is single variable  
\begin{equation}
\dfrac{\partial p}{\partial t} = - \dfrac{\partial }{\partial \rho} \left\{\left[ \rho + \left(D^2-1 \right) \rho^3 \right] p \right\} + \dfrac{D^2}{2} \dfrac{\partial^2 }{\partial \rho^2} \left( \rho^4 p \right) \label{sec4-eq3}
\end{equation}
The stationary distribution can be found analytically
\begin{equation}
p_\up{st}(\rho) = \mathcal{N} \rho^{-2\left(1+\frac{1}{D^2}\right)} \exp\left( -\dfrac{1}{D^2 \rho^2}\right) \label{sec4-eq4}
\end{equation}
where $\mathcal{N}$ is a constant determined through a normalization condition $\int_0^{+\infty} p_\up{st}(\rho) d\rho =1$. It is worth noticing that the same result cannot be obtained if the original problem is considered, because in the SDE \eqref{sec4-eq1} the amplitude equation and the Ornstein-Uhlenbeck process are coupled.

\begin{figure}[tb]
	\centering
	\includegraphics[angle=-90,width=45mm]{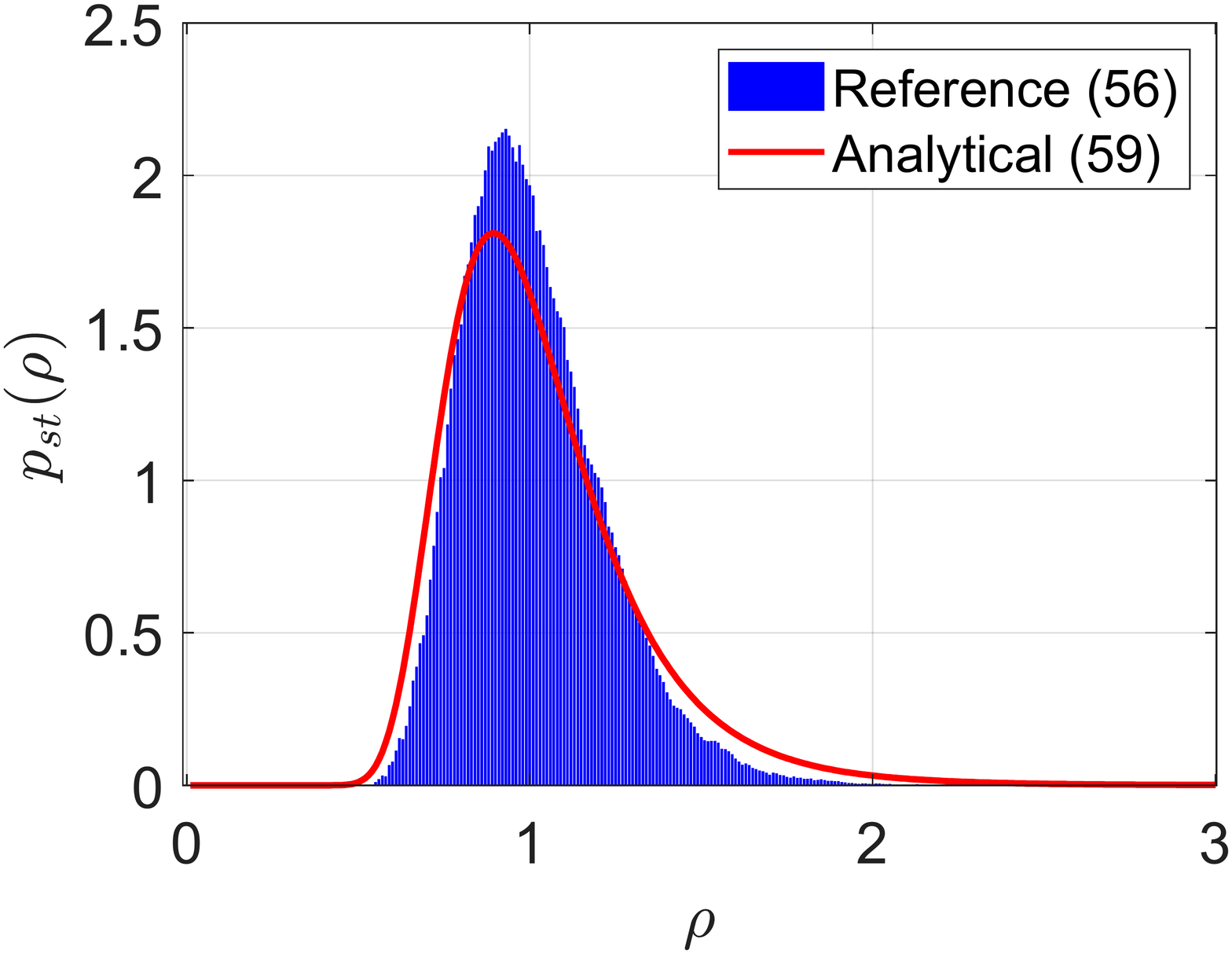}%
	\includegraphics[angle=-90,width=45mm]{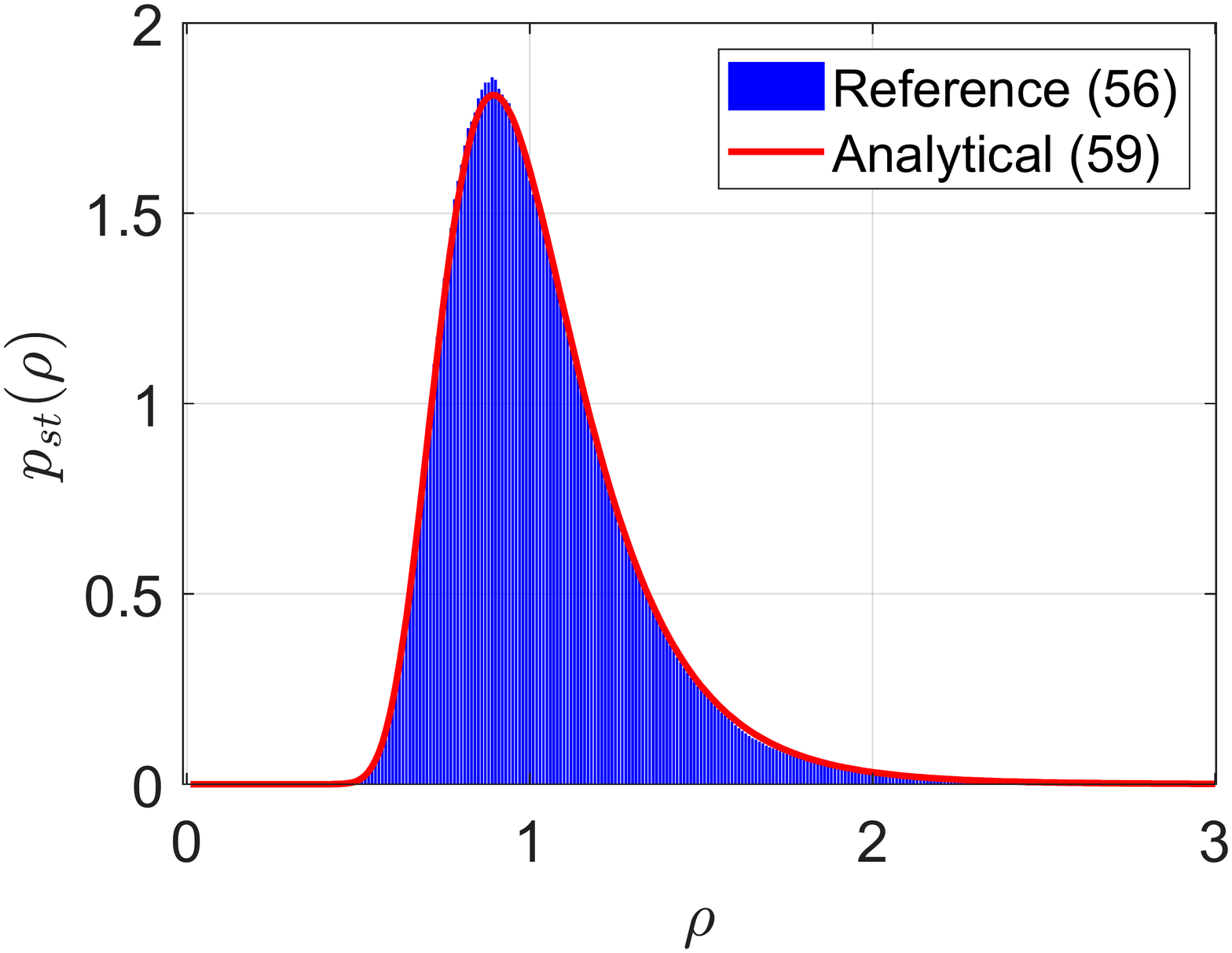}%
	\caption{Stationary distribution for the amplitude of a Stuart-Landau oscillator, for different values of the correlation time $\tau$. Blue bars are obtained from a numerical solution of the oscillator subject to colored noise \eqref{sec4-eq1}. The red line is the theoretical stationary distribution \eqref{sec4-eq4} obtained with the white noise approximation. On the left the results for $\tau=0.25$, on the right for $\tau=0.1$. Other parameters are $\alpha=4$, $\beta=2$, $D=0.5$.\label{figure1}}
\end{figure}

The theoretical prediction \eqref{sec4-eq4} is compared to the amplitude stationary distribution obtained through numerical integration of \eqref{sec4-eq1} in figure \ref{figure1}. Milstein numerical integration scheme has been used in the simulation, and the probability to find the amplitude in the interval $\rho + d \rho$ has been evaluated as the fraction of time spent in that interval divided by the total simulation length. As expected, the accuracy of the white noise approximation increases as the noise correlation time $\tau$ decreases.

In absence of noise the Stuart-Landau oscillator admits of an asymptotically stable limit cycle $\bx_s(t) = [(\alpha-\beta)t, 1]^T$. The Floquet vectors are $\bu_1(t) = [1,0]^T$, $\bu_2(t) =[\beta,1]^T$, while the co-vectors are $\bv_1(t) = [1,-\beta]^T$, $\bv_2(t) = [0,1]^T$. It is straightforward to derive the phase and amplitude deviation equations
\begin{align}
\nonumber d \theta = & \bigg\{ 1 + \dfrac{1}{\alpha - \beta} \bigg[ -\beta R + \left( \dfrac{D^2}{2} - \beta \right) (1+R)^2  \\[2ex]
\nonumber & - \beta \left( D^2 - 1 \right) (1+R)^3 \bigg] \bigg\} dt\\[2ex]
&  + \dfrac{D}{\alpha - \beta} (1+R)[1-\beta(1+R)] dW_t  \label{sec4-eq5a} \\[2ex]
d R = & \bigg[ 1 + R + \left( D^2-1 \right) (1+R)^3 \bigg] dt + D (1+R)^2 dW_t \label{sec4-eq5b}
\end{align} 

The linearized SDE for the amplitude is
\begin{equation}
d R = \left[ \left(- 2 + 3D^2 \right) R + D^2 \right] dt + D \, dW_t \label{sec4-eq6}
\end{equation}
and the stationary distribution obtained solving the associated Fokker-Planck equation is 
\begin{equation}
\hat p_\up{st}(R) =  \hat{\mathcal{N}} \,  \exp \left( \dfrac{3D^2-1}{D^2} R^2 + 2 R \right) \label{sec4-eq7}
\end{equation}
where $\hat{\mathcal{N}}$ is the normalization constant.

\begin{figure}[tb]
	\centering
	\includegraphics[angle=-90,width=70mm]{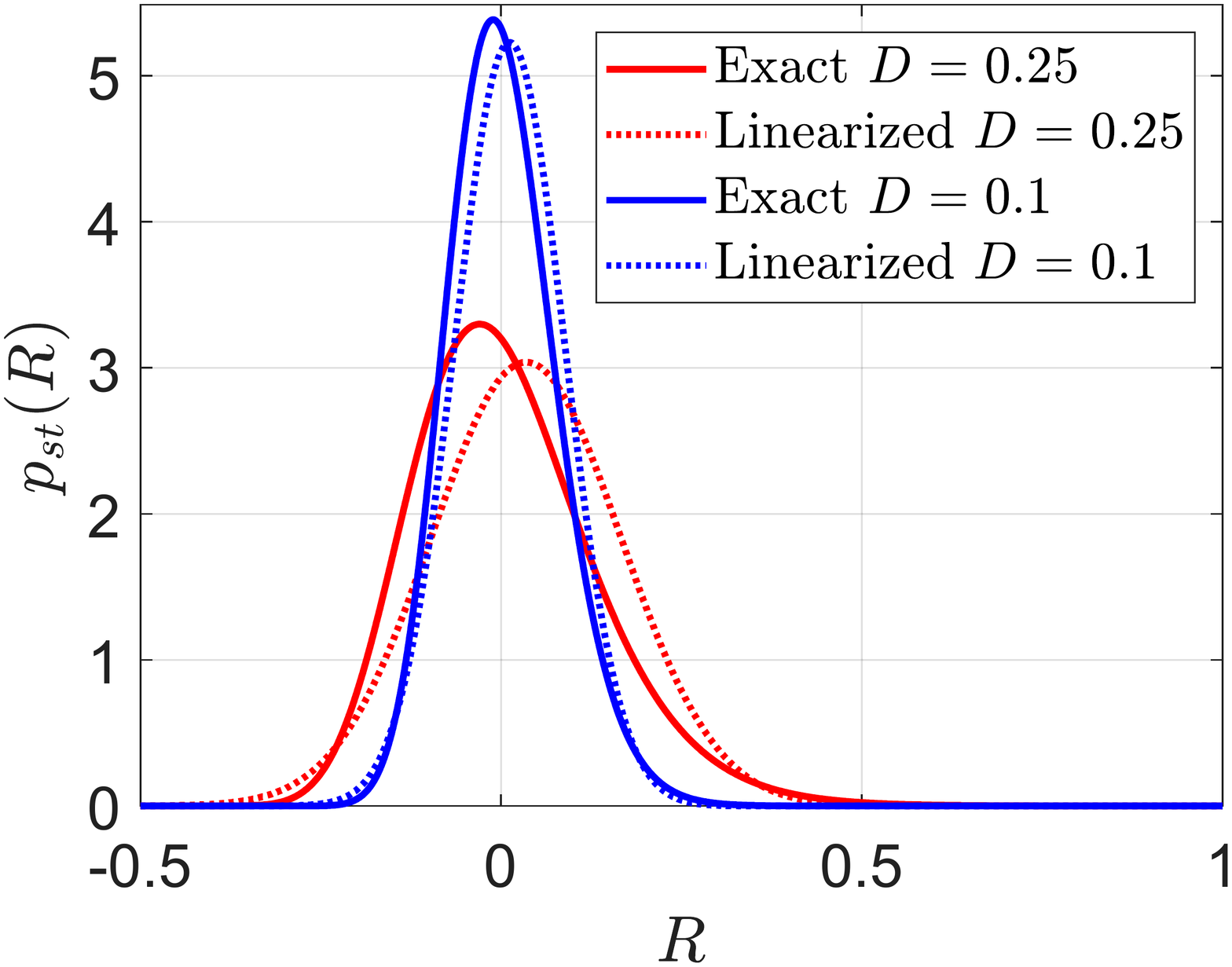}%
	\caption{Stationary distribution for the amplitude deviation of a Stuart-Landau oscillator, for different values of $D$. Solid and dashed blue lines are the stationary distributions \eqref{sec4-eq4} and \eqref{sec4-eq7}, respectively, for $D=0.25$. Solid and dashed red lines are the stationary distributions \eqref{sec4-eq4} and \eqref{sec4-eq7}, respectively, for $D=0.5$.\label{figure2}}
\end{figure}

\begin{figure}[tb]
	\centering
	\includegraphics[angle=-90,width=45mm]{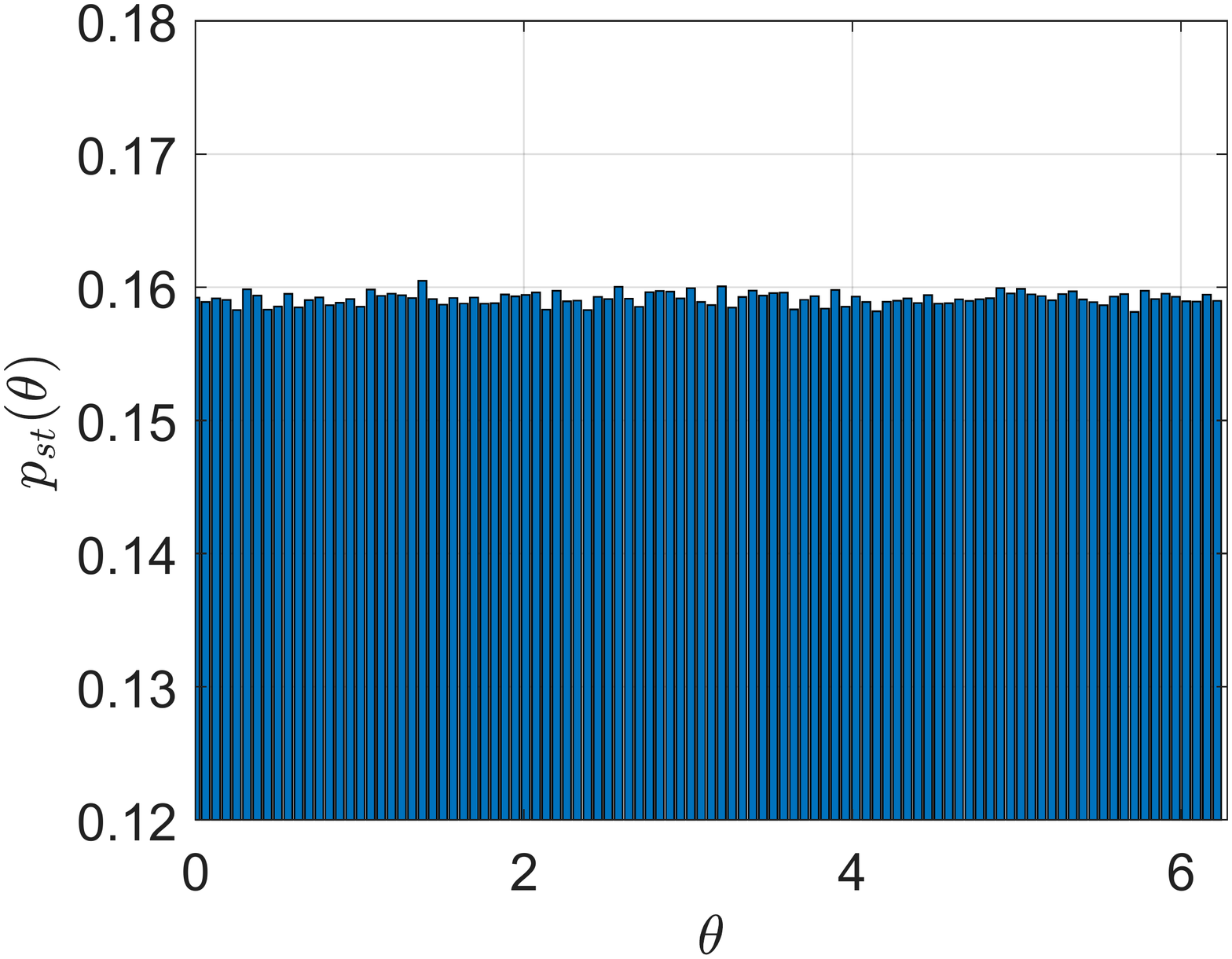}%
	\includegraphics[angle=-90,width=45mm]{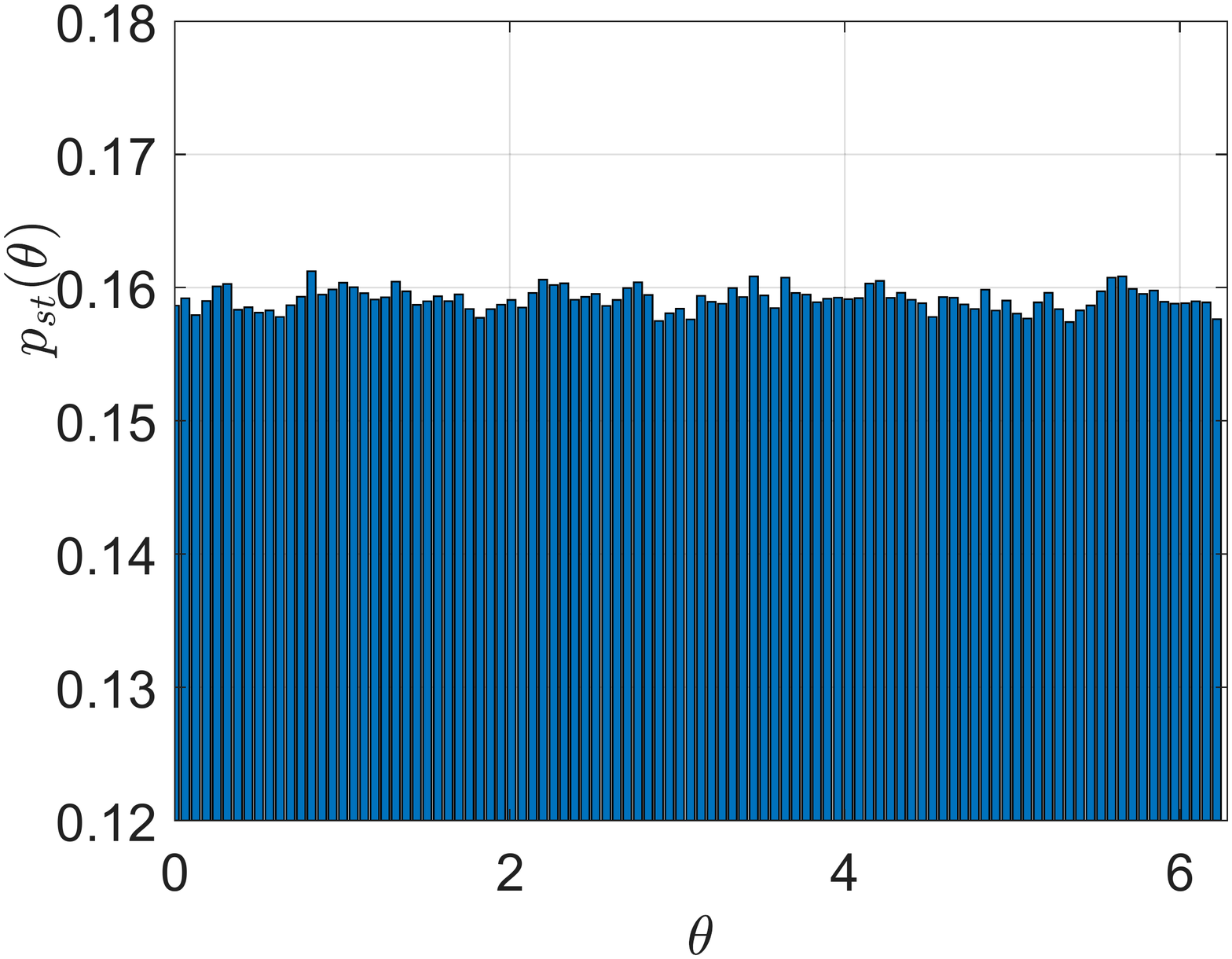}%
	\caption{Stationary distribution for the phase of a Stuart-Landau oscillator, for two different values of $D$. Left: $D=0.2$. Right $D=0.5$. Parameters are $\alpha=4$, $\beta=2$.\label{figure3}}
\end{figure}

Figure~\ref{figure2} shows the comparison between the amplitude deviation stationary distribution for the full system \eqref{sec4-eq4}, and its counterpart for the linearized system \eqref{sec4-eq7}, for different values of $D$. As expected, the distribution of the linearized system approximates well the full distribution for small values of $D$. The stationary distribution for the phase, obtained using numerical integration, is shown in figure~\ref{figure3} for two different values of $D$. It confirms that the uniform distribution is a good approximation even for fairly large values of $D$.  

Solving equations \eqref{sec3-eq25}, \eqref{sec3-eq26}, we find the first two moments
\begin{align}
\langle R \rangle = & \dfrac{D^2}{2-3D^2} \label{sec4-eq8a} \\[1ex]
\langle R^2 \rangle = & \dfrac{2D^2 \langle R \rangle + D^2}{4-6D^2} \label{sec4-eq8b}
\end{align}
Finally, taking stochastic expectation on both sides of the first of \eqref{sec4-eq5a} and neglecting $\mathcal{O}(R^3)$ terms we obtain the expected angular frequency
\begin{align}
\nonumber \left\langle \dfrac{d \theta}{dt} \right\rangle = & 1 + \dfrac{1}{\alpha - \beta} \bigg\{ \dfrac{D^2}{2}(1-2\beta) + D^2(1-3 \beta) \langle R \rangle \\[1ex]
& + \left[ \dfrac{D^2}{2} (1-3\beta) + 2 \beta \right] \langle R^2 \rangle \bigg\}  \label{sec4-eq9}
\end{align}

A comparison with \cite{demir2002} can be made making explicit \eqref{sec3-eq28} 
\begin{align}
d \theta = & \left( 1 + \dfrac{1-\beta}{\alpha - \beta} \eta_t \right) dt \label{sec4-eq9b}\\[1ex]
\tau d\eta_t = & - \eta_t \, dt + D \, dW_t \label{sec4-eq9c}
\end{align}
and taking the stochastic expectation on both sides of \eqref{sec4-eq9b}. Using \eqref{sec1-eq3} we find the expected angular frequency
\begin{equation}
\left \langle \dfrac{d \theta}{dt} \right \rangle = 1 + \eta_0 \, e^{-\frac{t}{\tau}} \label{sec4-eq9d}
\end{equation}
Therefore, according to the model in \cite{demir2002}, noise has no influence at all on the asymptotic expected angular frequency.

\begin{figure}[tb]
	\centering
	\includegraphics[angle=-90,width=70mm]{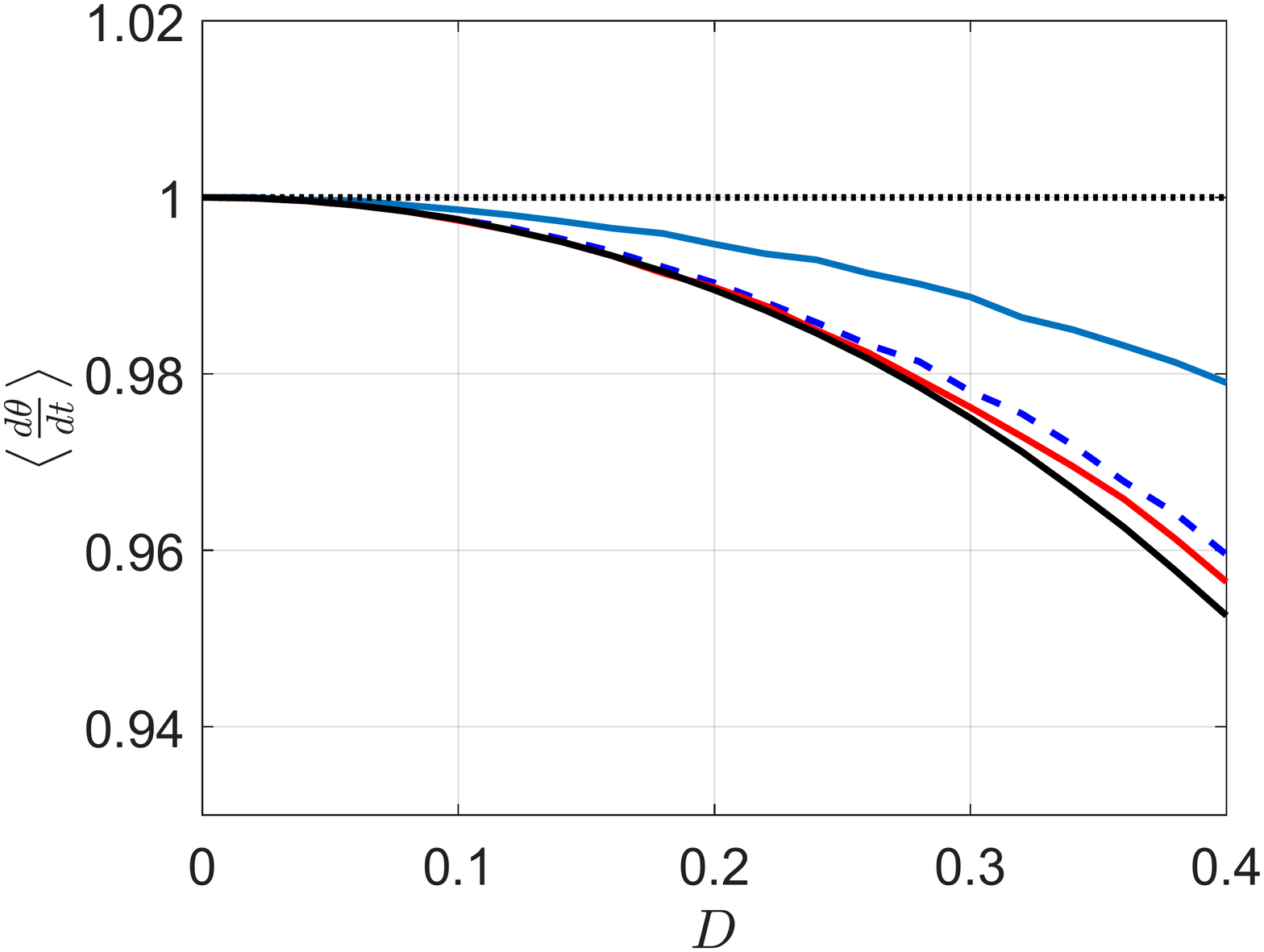}%
	\caption{Expected normalized angular frequency for a Stuart-Landau oscillator versus $D$. Blue lines: Numerical result for the system with colored noise \eqref{sec4-eq1}. Solid: $\tau=0.5$. Dashed: $\tau=0.1$. Red line: numerical result for the equivalent system with white Gaussian noise \eqref{sec4-eq2}. 
	Black dotted line: Asymptotic theoretical prediction \eqref{sec4-eq9d}. 
	Black solid line: Theoretical prediction \eqref{sec4-eq9}. Parameters are $\alpha=4$, $\beta=2$.\label{figure4}}
\end{figure}

Figure~\ref{figure4} shows the expected normalized angular frequency for the Stuart-Landau oscillator with colored noise \eqref{sec4-eq1}, the equivalent system with white noise \eqref{sec4-eq2}, and the theoretical predictions \eqref{sec4-eq9d} and \eqref{sec4-eq9}, as functions of $D$. Under the hypothesis that the system is ergodic, the normalized expected frequencies for systems \eqref{sec4-eq1} and \eqref{sec4-eq2} have been obtained through numerical integration, using the time average
\begin{equation}
\left \langle \dfrac{d \theta}{dt} \right\rangle = \dfrac{1}{\alpha-\beta} \, \dfrac{\phi(t_2) - \phi(t_1)}{t_2 - t_1} \label{sec4-eq10}
\end{equation} 
for $t_2\gg t_1$.

\begin{figure}[tb]
	\centering
	\includegraphics[angle=-90, width=60mm]{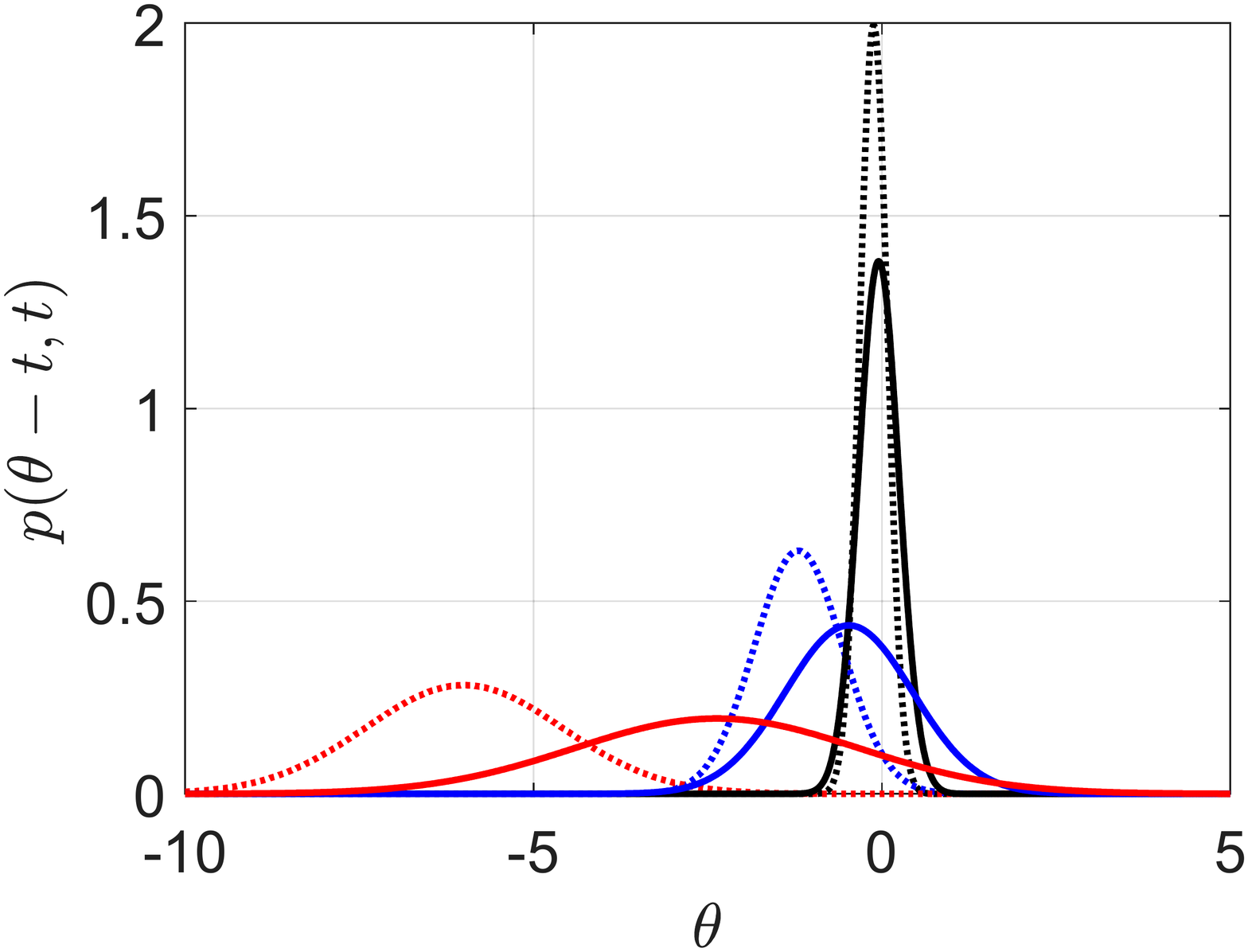}%
	\caption{PDF $p(\theta-t,t)$ at three different time instants. Solid lines: PDF with $\langle R \rangle$ and $\langle R^2 \rangle$  given by \eqref{sec4-eq8a} and \eqref{sec4-eq8b}, respectively. Dashed lines: PDF obtained neglecting amplitude fluctuations, that is imposing $\langle R \rangle = \langle R^2 \rangle= 0$. Other parameters are $\alpha=4$, $\beta=2$, $D=0.4$. \label{figure11}}
\end{figure}

An analysis of the phase equation \eqref{sec4-eq5a} provides further information. Averaging over the amplitude, substituting \eqref{sec4-eq8a} and \eqref{sec4-eq8b}, and neglecting $\mathcal{O}(R^3)$ terms, proves that the angle variable is well approximated by a Brownian motion with drift\footnote{This process is sometime referred to simply as Brownian motion, whereas the case $\mu=0$ is called Standard Brownian motion, in accordance with the corresponding PDFs.} $\theta(t) = \mu t + \sigma W_t$, where
	\begin{align}
	\mu = &  1 + \dfrac{1}{\alpha - \beta} \bigg\{ \dfrac{D^2}{2}(1-2\beta) + D^2(1-3 \beta) \langle R \rangle \\[1ex]
	& + \left[ \dfrac{D^2}{2} (1-3\beta) + 2 \beta \right] \langle R^2 \rangle \bigg\} \\[2ex]
	\sigma = & \dfrac{D}{\alpha-\beta} \left[ 1 - \beta + (1- 2\beta) \langle R \rangle - \beta \langle R^2 \rangle \right] 
	\end{align}
For the sake of simplicity we assume a perfectly localized initial condition $\theta(0)=0$, and that $\theta \in (-\infty,+\infty)$, with boundary conditions $p(\pm \infty,t) = 0$.  
Then the phase has a normal distribution $p(\theta,t) \sim N(\mu t, \sigma^2 t)$, and the auto-correlation is 
	\begin{equation}
	R(\theta(t),\theta(s)) = \sqrt{\dfrac{\min(t,s)}{\max(t,s)}}
	\end{equation} 
Figure \ref{figure11} shows the PDF $p(\theta-t,t)$ for the phase deviation (i.e. the difference between $\theta(t)$ and the phase in absence of noise) at three different time instants.

\begin{figure}[tb]
	\centering
	\includegraphics[angle=-90, width=65mm]{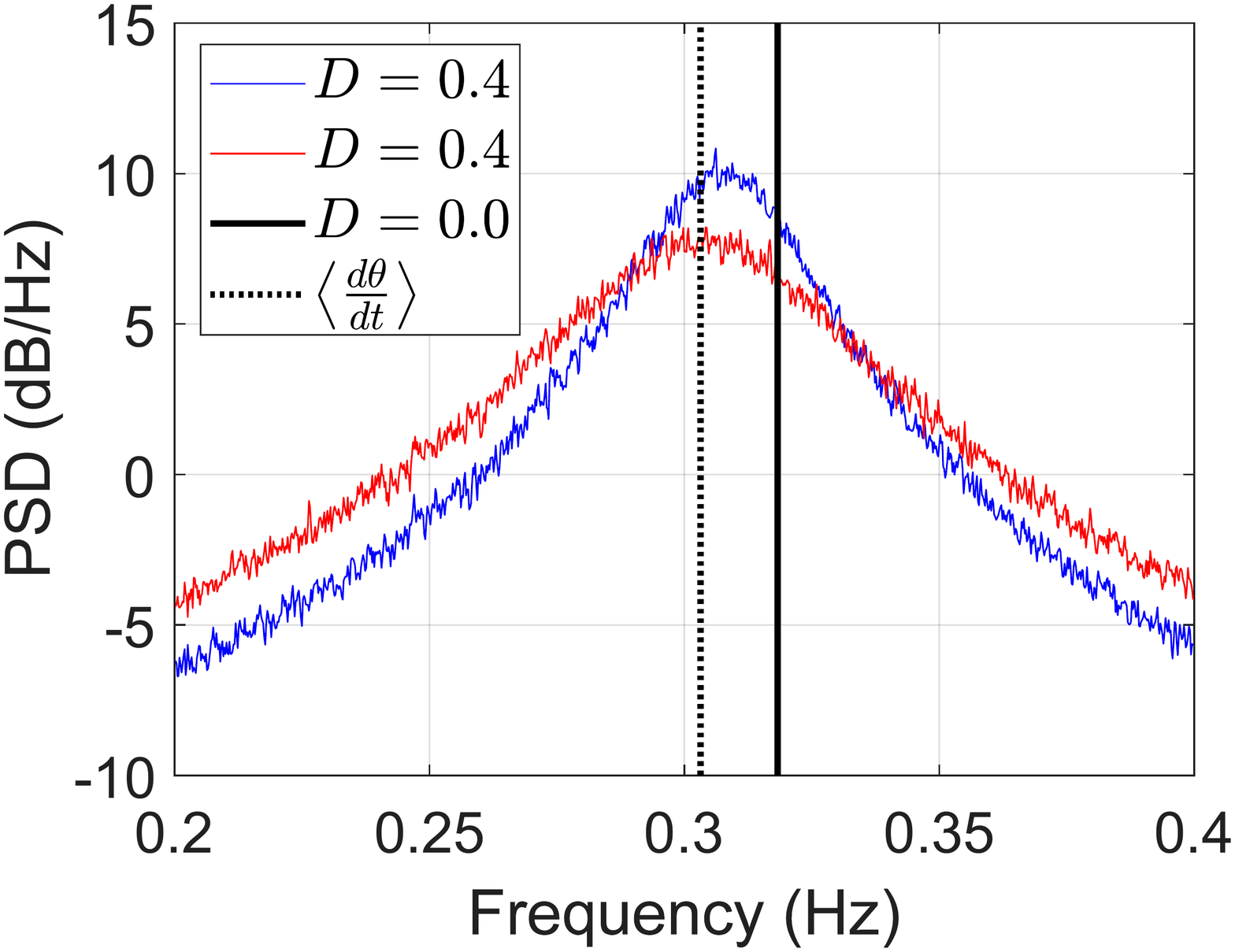}%
	\caption{Power spectral density for the orbital noise component $\bx_s$ of the Stuart--Landau oscillator computed using the Welch's method. Blue line: PSD for the full system with colored noise \eqref{sec4-eq1}, noise intensity $D=0.4$ and noise correlation time $\tau=0.5$. Black solid line: PSD for the full system without noise. Red line: PSD obtained from the reduced phase equation. Black dashed line: expected angular frequency $\left \langle \frac{d \theta}{dt} \right \rangle$ given by \eqref{sec4-eq9}. Parameters are $\alpha = 4$ and $\beta=2$. \label{figure12}}
\end{figure}

The power spectral density (PSD) for the orbital noise component, calculated using Welch's method together with an average over 200 realizations of the corresponding time-domain process, is shown in figure~\ref{figure12}. The blue line is the PSD for the orbital noise $x_s = \rho(t) \cos\phi(t)$ in presence of the colored noise source as defined in \eqref{sec4-eq1}. The noise correlation time is $\tau = 0.5$ and the noise intensity is $D=0.4$. The red line is the PSD for the orbital noise component $x_s = (1+\mu) \cos[(\alpha-\beta)\theta(t) + \beta \mu]$,  where $\theta$ is the solution of the reduced phase equation \eqref{sec3-eq27}, again for $D=0.4$. The two PSDs show excellent agreement around the carrier frequency, while the system with colored noise shows a significantly reduced power content at high frequency (not shown in this figure). The black solid line is the PSD for the system without noise, here shown to put in evidence the frequency shift induced by noise. PSDs were obtained considering $10^4$ oscillations, divided into $2^{28}$ points, corresponding to a sampling rate of 5369 samples per second for the discrete Fourier transform calculation. Finally, the black dotted line marks the theoretical prediction for the expected angular frequency $\left \langle {d \theta}/{dt} \right \rangle$ as given by \eqref{sec4-eq9}.

\subsection{van der Pol oscillator with colored noise}

\begin{figure}[tb]
	\centering
	\includegraphics[angle=0, width=60mm]{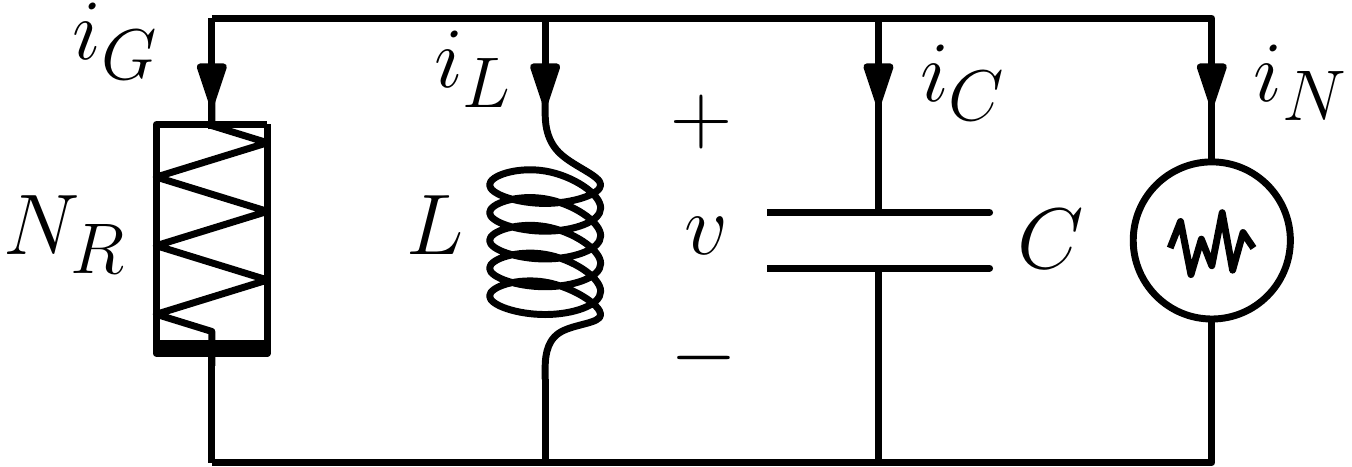}%
	\caption{Second order nonlinear oscillator. \label{figure10}}
\end{figure}

As a second example we consider the nonlinear oscillator (van der Pol) shown in figure \ref{figure10}. The nonlinear resistor $N_R$ is assumed to be noiseless, with characteristic $i_G = g(v) = {v^3}/{3} - v$. The random source on the right models environment and internal noise.
Using Kirchhoff current law it is straightforward to derive the state equations
\begin{align}
d v = & u \, dt' \\
d u = & \left[ -\dfrac{1}{LC} v - \dfrac{1}{C} g'(v) u - \dfrac{1}{C} s_n(t') \right] dt' 
\end{align}
where $i_N$ is the integral of the stochastic process $s_n$. With the change of variables
\begin{equation*}
t =  \dfrac{1}{\sqrt{LC}}\, t'\qquad x_1 = v \qquad x_2 = \sqrt{LC}\, u
\end{equation*}
and assuming that the random source $s_n$ is a colored noise modulated by the current through the capacitor
\[ s_n(t) = \sqrt{\frac{C}{L}} \,  u \, \eta_{t}\]
we obtain the state equations 
\begin{equation}
\begin{array}{rl}
d x_1 = & x_2 \, dt \\[1ex]
d x_2 = & \left[ - x_1 + \alpha\left( 1-x_1^2 \right) x_2 + x_2 \eta_t \right] dt \\[1ex]
\tau d \eta_t = & - \eta_t \, dt + D \, dW_t   
\end{array} \label{sec4-eq11}
\end{equation}
where $\alpha = \sqrt{{L}/{C}}$.

Transformation to the equivalent system with white noise yields
\begin{equation}
\begin{array}{rl}
d x_1 = & x_2 \, dt \\[1ex]
dx_2 = & \left[ - x_1 + \alpha \left( 1 - x_1^2 \right) x_2 + \dfrac{D^2}{2} x_2 \right] dt + D \, x_2 \, d W_t
\end{array}\label{sec4-eq12}
\end{equation}

\begin{figure}[tb]
	\centering
	\includegraphics[angle=-90, width=80mm]{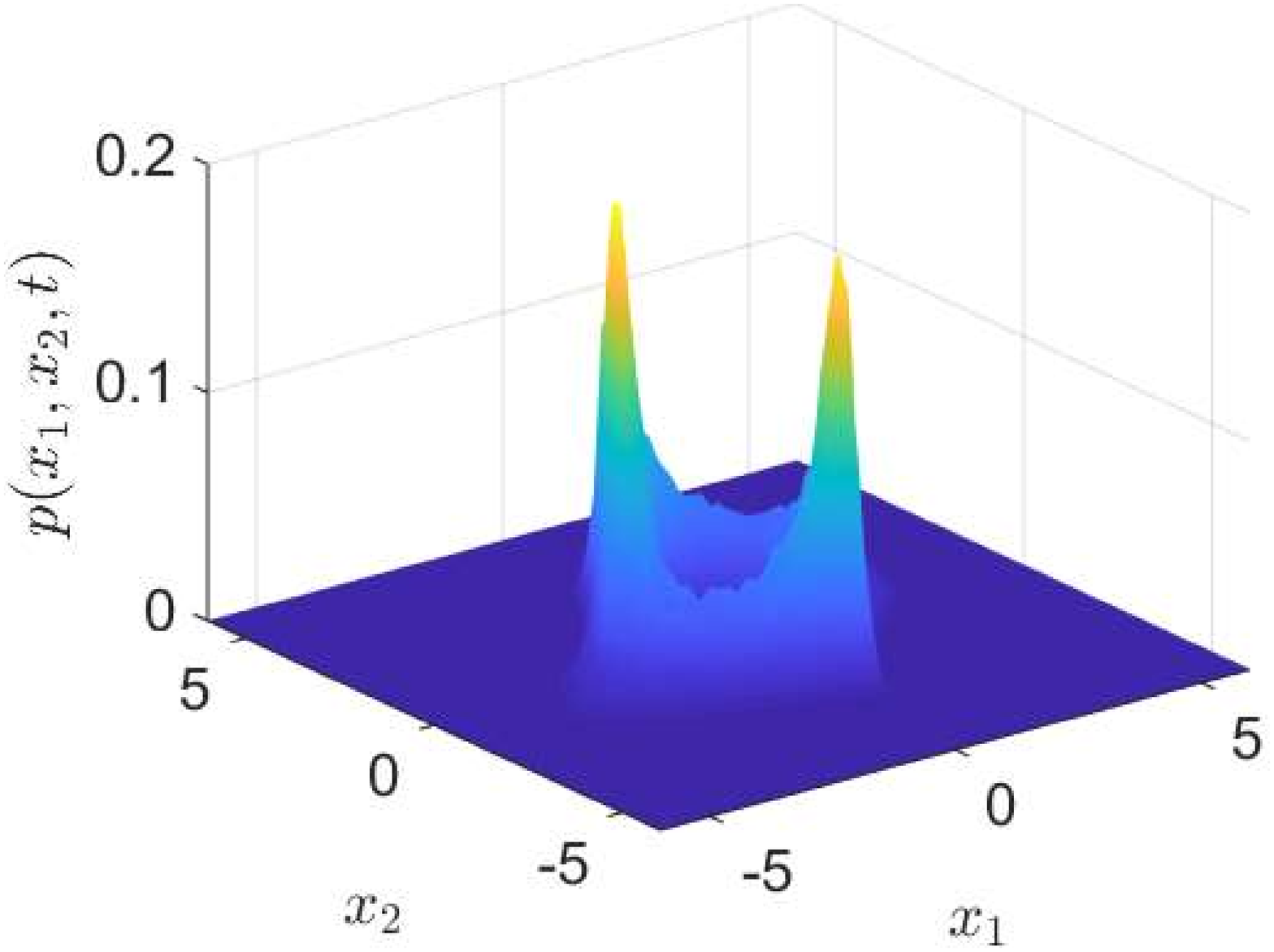}%
	\caption{Probability density function for the van der Pol oscillator with colored noise. Parameters are $\alpha=0.5$, $D=0.5$. Noise correlation time is $\tau=0.5$. \label{figure5}}
\end{figure}

\begin{figure}[tb]
	\centering
	\includegraphics[angle=-90, width=80mm]{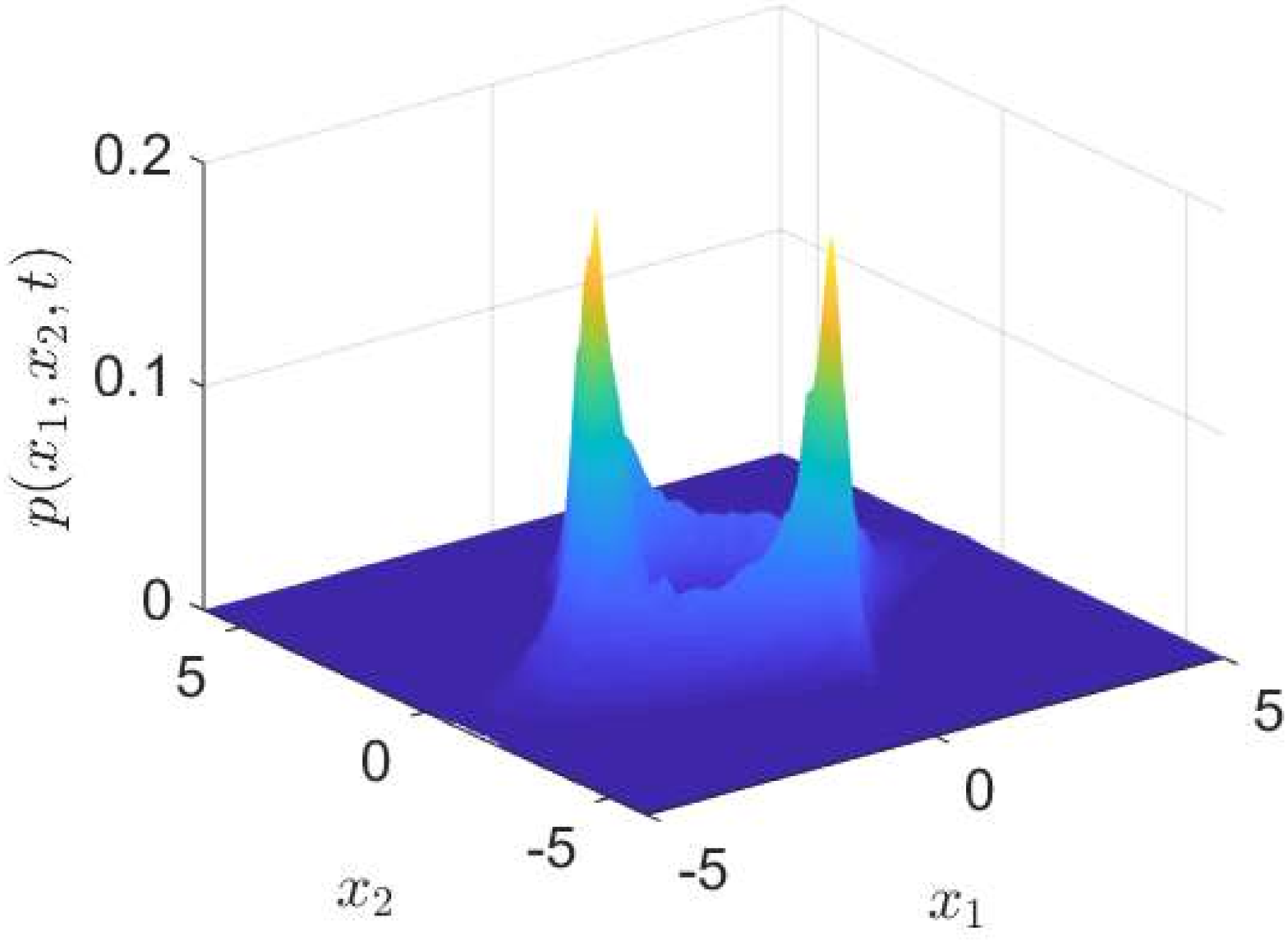}%
	\caption{Probability density function for the van der Pol oscillator with colored noise. Parameters are $\alpha=0.5$, $D=0.5$. Noise correlation time is $\tau=0.1$.\label{figure6}}
\end{figure}

\begin{figure}[tb]
	\centering
	\includegraphics[angle=-90, width=80mm]{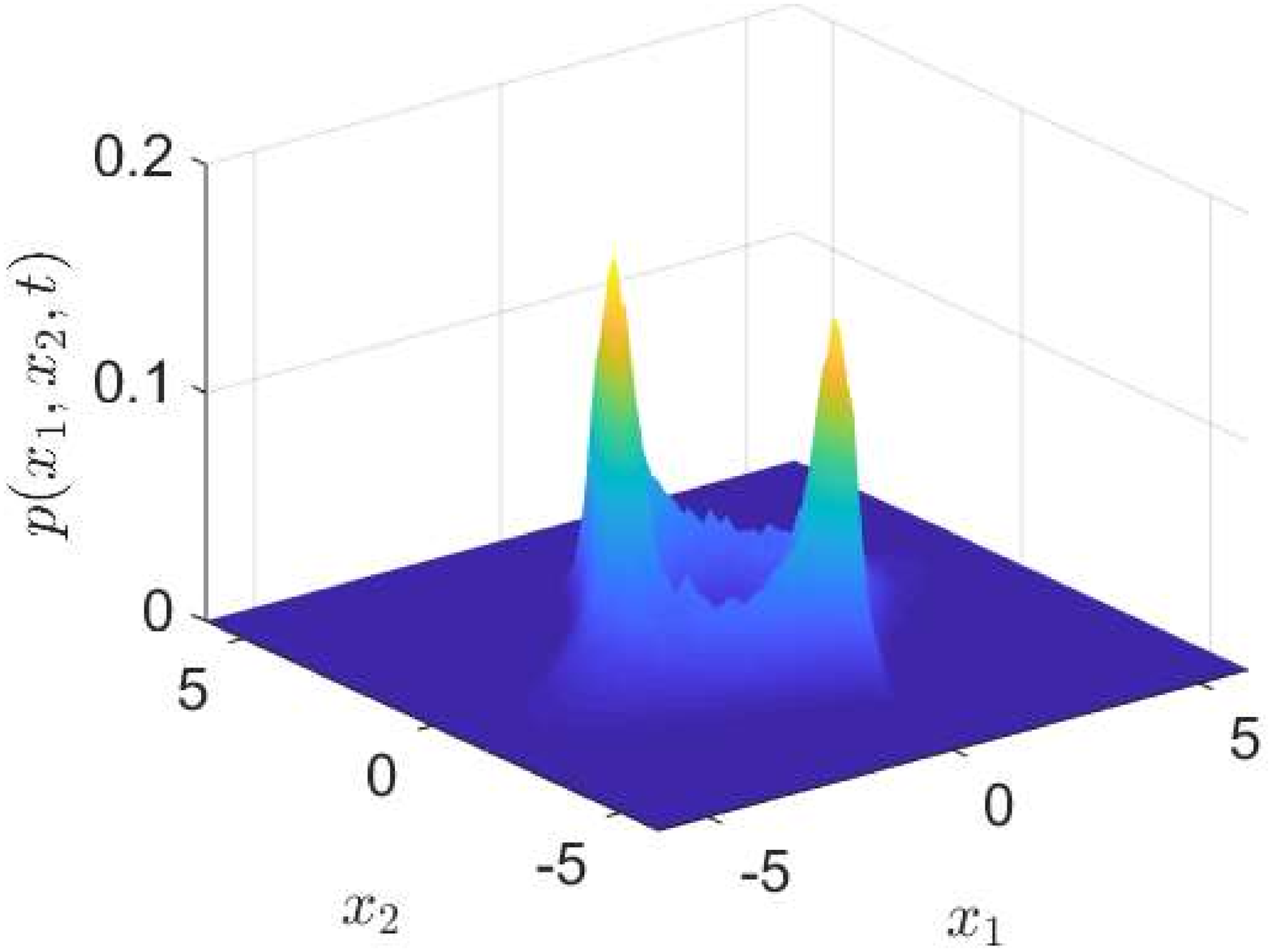}%
	\caption{Probability density function for the van der Pol oscillator with white Gaussian noise. Parameters are $\alpha=0.5$, $D=0.5$.\label{figure7}}
\end{figure}

Figures~\ref{figure5} and \ref{figure6} show the PDF $p(x_1,x_2,t)$,  at the same time instant, for the van der Pol oscillator with colored noise \eqref{sec4-eq11} and noise correlation time $\tau=0.5$ and $\tau=0.1$, respectively. Figure \ref{figure7} show the PDF for the equivalent system with white Gaussian noise \eqref{sec4-eq12}. The PDF has been computed from numerical simulations, the same initial condition has been used in all cases\footnote{Parameter values are conveniently chosen to simplify calculations and to highlight higher order contributions of noise.}. 

\begin{figure}[tb]
	\centering
	\includegraphics[angle=-90, width=70mm]{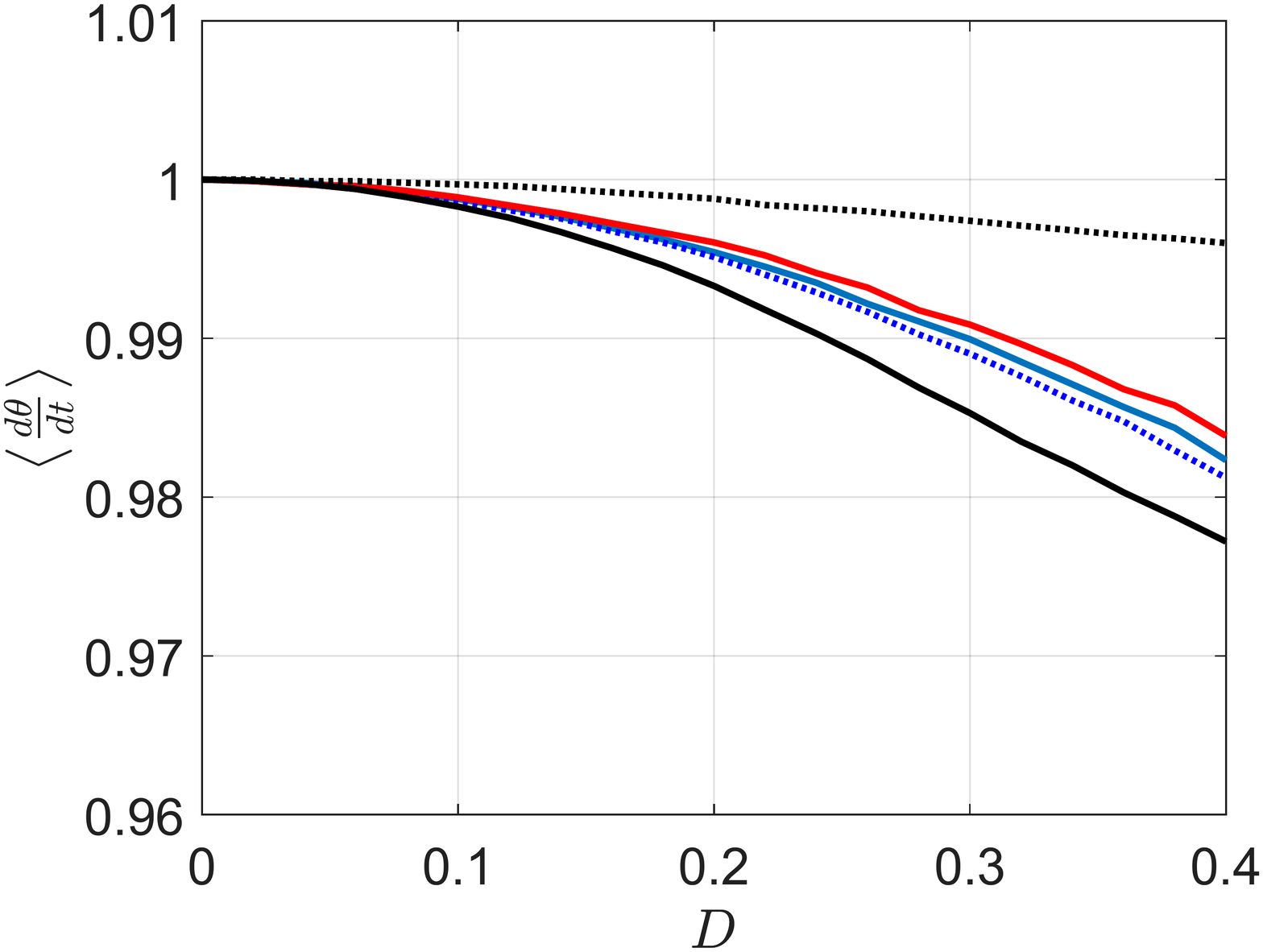}%
	\caption{Expected normalized frequency for a van der Pol oscillator versus $D$. Blue lines: Numerical result for the system with colored noise \eqref{sec4-eq11}. Solid: $\tau=0.5$. Dotted: $\tau=0.1$. Red line: numerical result for the equivalent system with white Gaussian noise \eqref{sec4-eq12}. {
	Black dotted line: expected  frequency obtained through numerical integration of \eqref{sec3-eq28}, \eqref{sec3-eq29}.} 
	Black solid line: Theoretical prediction using the proposed method. Parameter $\alpha=0.5$.\label{figure8}}
\end{figure}

Because the limit cycle and the Floquet vectors cannot be found analytically for the van der Pol oscillator, we resort to semi analytical techniques and numerical methods for the analysis. In particular, we have developed a methodology based on the following steps:
\begin{itemize}
	\item First, the limit cycle of the noiseless system is determined in a semi analytical form using the Harmonic Balance technique \cite{bonnin2008,traversa2008}.
	\item The semi analytical expression of the limit cycle is used to determine the Floquet vectors and co-vectors. Because the example under investigation is a second order system, Floquet vectors an co-vectors are computed using the formulas given in \cite{bonnin2012}. For higher order systems, they can be found exploiting efficient numerical techniques \cite{traversa2008,IET,AEU,TCAD}.
	\item The limit cycle, and the Floquet vectors and co-vectors are used to determine the functions in equations \eqref{sec3-eq7}-\eqref{sec3-eq16}.
	\item The functions $\boldsymbol{M}$, $\boldsymbol{m}$, $\boldsymbol{N}$ and $\boldsymbol n$ given by \eqref{sec3-eq20}-\eqref{sec3-eq23} are calculated, using the uniform distribution $p_\up{st}(\theta) = 1/ 2\pi$ for averaging. Equations \eqref{sec3-eq25} and \eqref{sec3-eq26} are solved to find $\langle R_i \rangle$ and $\langle R_i R_j \rangle$ for all $i,j$.  
	\item The reduced phase equation is written  and can be analyzed to determine the expected normalized angular frequency, frequency shift, diffusion constant and so on. 
\end{itemize}
Figure~\ref{figure8} shows the expected normalized angular frequency for the van der Pol system with colored noise \eqref{sec4-eq11}, as a function of $D$. The expected frequency exhibits little dependence on the noise correlation time, in fact the blue curves (the solid curve corresponds to $\tau=0.5$, while the dashed one to  $\tau=0.1$) are very close. The red line represents the expected normalized angular frequency for the equivalent system with white Gaussian noise \eqref{sec4-eq12}. The curve has been determined through numerical integration of the full phase-amplitude deviation SDEs. Finally the black curves represents the theoretical predictions. The dotted line is the expected angular frequency determined through numerical integration of \eqref{sec3-eq28}, \eqref{sec3-eq29} \cite{demir2002}. The solid line is the theoretical prediction given by our phase reduced model, obtained using the methods described above. 

\begin{figure}[tb]
	\centering
	\includegraphics[angle=-90, width=45mm]{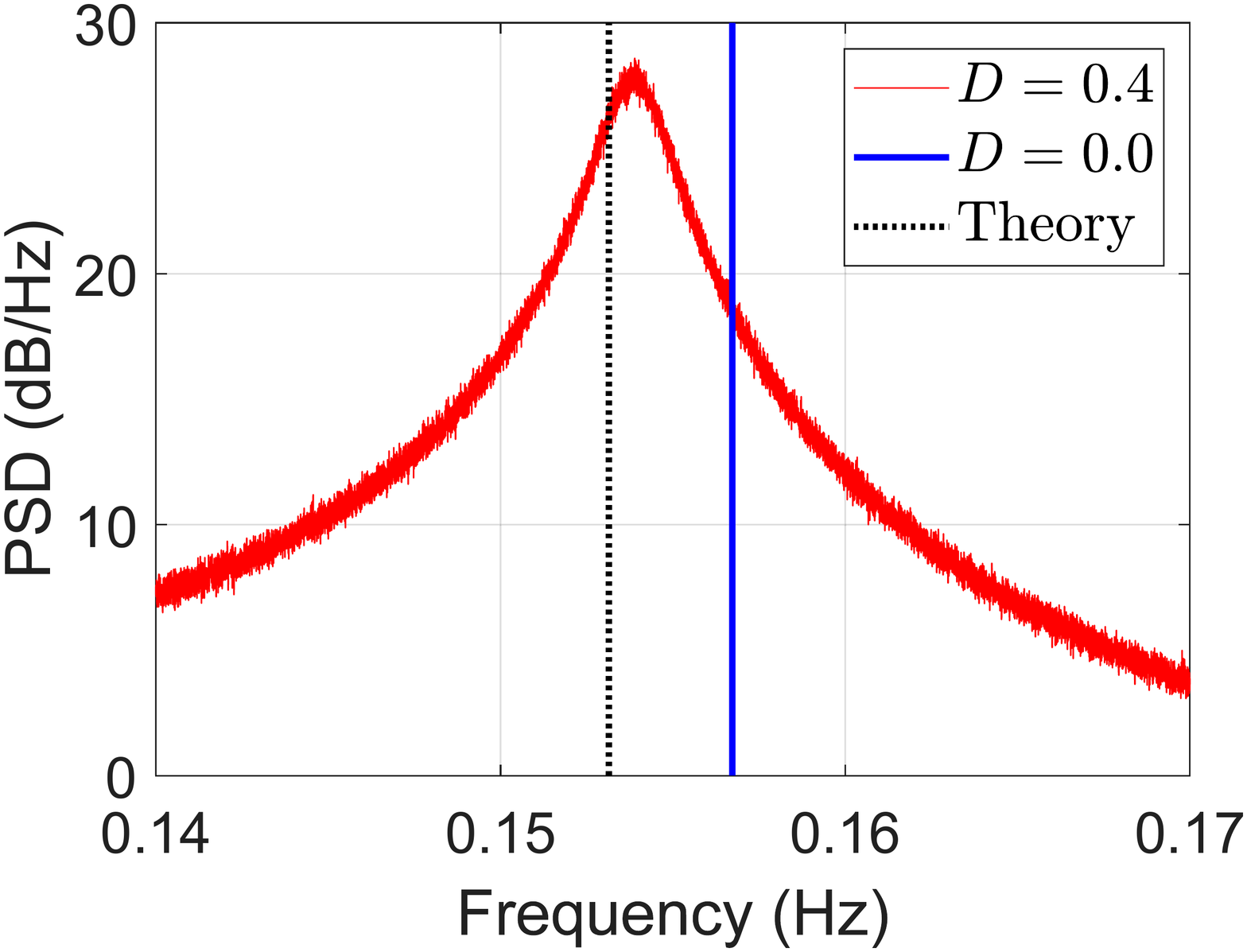}%
	\includegraphics[angle=-90, width=45mm]{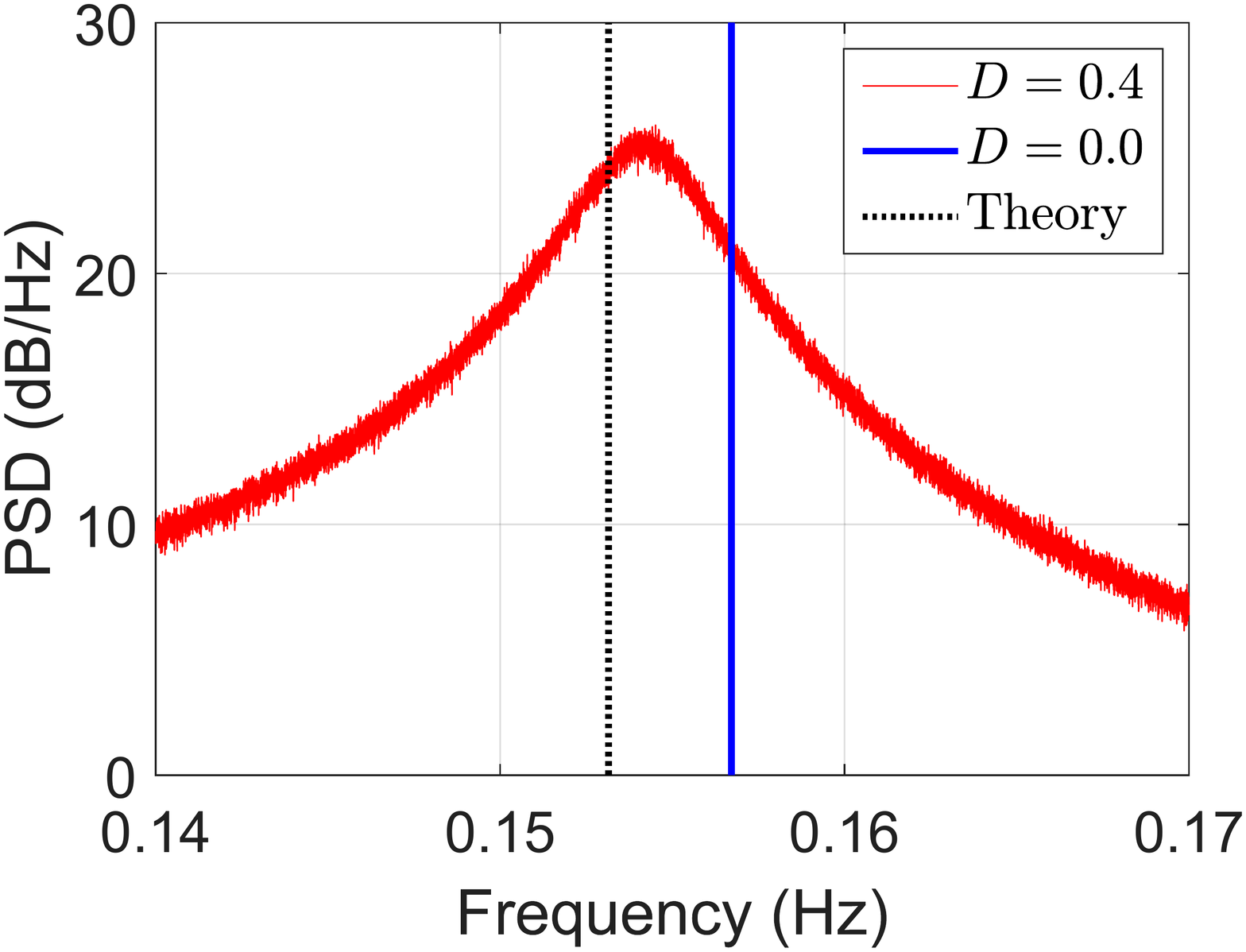}%
	\caption{Power Spectral Density for numerical solution of a van der Pol oscillator computed using Welch's algorithm. Left: PSD for the system with colored noise \eqref{sec4-eq11}, $\tau=0.5$. Right: PSD for the equivalent system with white Gaussian noise \eqref{sec4-eq12}. The black dashed line identifies the expected frequency obtained using our theoretical model. Parameter $\alpha=0.5$.\label{figure9}}
\end{figure}

As a further confirmation, we have computed the power spectral density for the numerical solution of the SDEs \eqref{sec4-eq11} and \eqref{sec4-eq12}. Figure~\ref{figure9} shows  the power spectral density calculated combining Welch's method with an averaging over 200 realizations to the numerical solution of the van der Pol system with colored and white noise, respectively. The discrete Fourier transform was calculated on a signal with time length $10^4 T_0$ ($T_0$ being the period of the noiseless oscillator determined exploiting the harmonic balance technique) and a sampling rate of $4207$ samples per second. The blue line is the PSD in the absence of noise, while the red line is PSD for the system with noise ($D=0.4$). Noise clearly shifts the position of the power peak. The black dashed line identifies the expected frequency found using our theoretical model.

\section{Conclusions}

In this paper we have investigated the effect of colored noise on phase noise in nonlinear oscillators. We used a general method that transforms nonlinear systems subject to colored noise, modeled as an Ornstein-Uhlenbeck process, into equivalent systems subject to white Gaussian noise. The original system has unmodulated (additive) noise, and as such it is free of the It\^o-Stratonovich dilemma. The transformation leads to two equivalent stochastic differential equations for the It\^o and the Stratonovich interpretations. Since the two equations are equivalent, they have the same solution and therefore which one should be used is just a matter of personal preference. An alternative description in terms of stochastic differential equations for the phase and the amplitude deviation is derived for the transformed system. The phase variable used coincide locally, in the neighborhood of the unperturbed limit cycle, with the asymptotic phase defined using the concept of isochrons. Using stochastic averaging, a reduced phase model is derived, where the system dynamics is described only in terms of the phase variable.  

The reduced phase model describes phase noise as a drift-diffusion process. The shift in the expected frequency is related to the variance of the colored noise and noise correlation time. A numerical procedure is presented for the solution of the phase equation, that provides more accurate results than other previously proposed models.

\bibliography{colored_noise_final}
\bibliographystyle{ieeetran}

\end{document}